\documentclass[12pt]{article}

\usepackage{amssymb,amsfonts,amsthm}%,psfig}
\usepackage{graphicx}
\usepackage{subfigure}
\usepackage{amsmath}
\usepackage{bm}             %% boldmath-command: \bm{}
\usepackage[mathscr]{eucal}

\usepackage{cancel}

\usepackage{psfrag}

\usepackage{multirow}

\usepackage{wasysym}

\usepackage{color}

\newcommand{\la}{\langle}
\newcommand{\ra}{\rangle}
\def\hsp5{\hspace{5mm}}

\newcommand{\sfrac}[2]{{\textstyle{#1\over#2}}}

\theoremstyle{plain}
\newtheorem{theorem}{Theorem}[section]

\theoremstyle{remark}

\title{\sc Global dynamics and asymptotics for monomial scalar field potentials and perfect fluids}
\begin{document}

\author{
\sc Artur Alho,$^{1}$\thanks{Electronic address:{\tt
aalho@math.ist.utl.pt}}\, Juliette Hell$^{2}$\thanks{Electronic address:{\tt
blanca@math.fu-berlin.de}}\, and Claes Uggla$^{3}$\thanks{Electronic
address:{\tt claes.uggla@kau.se}}\\
$^{1}${\small\em Center for Mathematical Analysis, Geometry and Dynamical Systems,}\\
{\small\em Instituto Superior T\'ecnico, Universidade de Lisboa,}\\ 
{\small\em Av. Rovisco Pais, 1049-001 Lisboa, Portugal.}\\
$^{2}${\small\em Freie Universit\"at Berlin, Institut f\"ur Mathematik,}\\
{\small\em Arnimallee 3D - 14195 Berlin, Germany.}\\
$^{3}${\small\em Department of Physics, University of Karlstad,}\\
{\small\em S-65188 Karlstad, Sweden.}}

%%%%%%%%%%%%%%%%%%%%%%%%%%%%%%%%%%%%%%%%%%%%%%%%%%%%%%%%%%%%%%%%%%%
%\date{\normalsize{February 27, 2014}}
\date{}
%%%%%%%%%%%%%%%%%%%%%%%%%%%%%%%%%%%%%%%%%%%%%%%%%%%%%%%%%%%%%%%%%%%
\maketitle
%%%%%%%%%%%%%%%%%%%%%%%%%%%%%%%%%%%%%%%%%%%%%%%%%%%%%%%%%%%%%%%%%%%

%%%%%%%%%%%%%%%%%%%%%%%%%%%%%%%%%%%%%%%%%%%%%%%%%%%%%%%%%%%%%%%%%%%
\begin{abstract}
%%%%%%%%%%%%%%%%%%%%%%%%%%%%%%%%%%%%%%%%%%%%%%%%%%%%%%%%%%%%%%%%%%%

We consider a minimally coupled scalar field with a monomial potential and a
perfect fluid in flat FLRW cosmology. We apply local and global dynamical
systems techniques to a new three-dimensional dynamical systems reformulation
of the field equations on a compact state space. This leads to a visual
global description of the solution space and asymptotic behavior. At late
times we employ averaging techniques to prove statements about how the
relationship between the equation of state of the fluid and the monomial
exponent of the scalar field affects asymptotic source dominance and
asymptotic manifest self-similarity breaking. We also situate the `attractor'
solution in the three-dimensional state space and show that it corresponds to
the one-dimensional unstable center manifold of a de Sitter fixed point,
located on an unphysical boundary associated with the dynamics at early
times. By deriving a center manifold expansion we obtain approximate
expressions for the attractor solution. We subsequently improve the accuracy
and range of the approximation by means of Pad{\'e} approximants and compare
with the slow-roll approximation.

%%%%%%%%%%%%%%%%%%%%%%%%%%%%%%%%%%%%%%%%%%%%%%%%%%%%%%%%%%%%%%%%%%
\end{abstract}
%%%%%%%%%%%%%%%%%%%%%%%%%%%%%%%%%%%%%%%%%%%%%%%%%%%%%%%%%%%%%%%%%%

%\centerline{\bigskip\noindent PACS numbers: 04.20.-q, 98.80.-k, 98.80.Bp,
%98.80.Jk}

%%%%%%%%%%%%%%%%%%%%%%%%%%%%%%%%%%%%%%%%%%%%%%%%%%%%%%%%%%%%%%%%%%
\section{Introduction}
%%%%%%%%%%%%%%%%%%%%%%%%%%%%%%%%%%%%%%%%%%%%%%%%%%%%%%%%%%%%%%%%%%

The present paper investigates general relativistic flat
Friedmann-Lema\^{i}tre-Robertson-Walker (FLRW) models with a minimally
coupled scalar field with a monomial potential, $V(\phi) =
\frac{1}{2n}(\lambda \phi)^{2n}$ ($\lambda>0$, $n=1,2,3,\dots$), and a
perfect fluid. The perfect fluid is assumed to obey a linear equation of
state, $p_m =(\gamma_m-1)\rho_m$, where $p_m$ and $\rho_m \geq 0$ are the
pressure and the energy density, respectively. The adiabatic index is assumed
to satisfy $0 <\gamma_m < 2$, where $\gamma_m = 1$ corresponds to dust and
$\gamma_m= 4/3$ to radiation. When $\gamma_m = 2/3$ the matter term $\rho_m$
can be reinterpreted as $-\frac12{}^3\!R$, where ${}^3\!R$ is the spatial
3-curvature of the open FLRW model, i.e., $\gamma_m = 2/3$ leads to equations
that are the same as those for a scalar field in open FLRW cosmology. The
case $\gamma_m=0$ corresponds to a matter content described by a cosmological
constant, i.e., $\rho_m=\Lambda$, while $\gamma_m = 2$ describes a stiff
perfect fluid; both cases are associated with significant bifurcations, and
we therefore refrain from discussing them.

The Einstein and matter field equations for these models are given by
\begin{subequations}\label{Hphieq}
\begin{align}
3H^2 &= \sfrac12 \dot{\phi}^2 + \frac{1}{2n}(\lambda\phi)^{2n} + \rho_m = \rho_\phi + \rho_m,\label{Gauss1}\\
\dot{H} &= -\sfrac{1}{2}\left(\dot{\phi}^2 + \gamma_m\rho_m\right),\label{Ray1}\\
0 &= \ddot{\phi} + 3H\dot{\phi} + \lambda^{2n}\phi^{2n-1},\label{KG}\\
\dot{\rho}_m &= - 3H\gamma_m\rho_m .\label{rhoeq}
\end{align}
\end{subequations}
Here an overdot signifies the derivative with respect to synchronous proper
time, $t$; $H$ is the Hubble variable, which is given by $H = \dot{a}/a$,
where $a(t)$ is the cosmological scale factor, and throughout we assume an
expanding Universe, i.e. $H>0$, where $H$ is related to the expansion
$\theta$ according to $H = \theta/3$. We use (reduced Planck) units such that
$c=1=8\pi G$, where $c$ is the speed of light and $G$ is the gravitational
constant (in the inflationary literature the gravitational constant $G$ is
often replaced by the Planck mass, $G=1/m_\mathrm{Pl}^2$).

Heuristically eq.~\eqref{KG} can be viewed as an equation for an anharmonic
oscillator with a friction term $3H\dot{\phi}$. This suggests that
$(\dot{\phi},\phi) \rightarrow (0,0)$ toward the future in an oscillatory
manner, which is indeed correct. This qualitative picture, however, does not
show how this comes about in a quantitative way, nor how the fluid affects
the situation via its influence on $H$. Running the time backwards allows one
to heuristically interpret $3H\dot{\phi}$ as an energy input, which suggests
that the scalar field oscillates with increasing amplitude toward the past,
but this picture breaks down in the limit $H\rightarrow \infty$. Even though
this is beyond the Planck regime, this limit is also needed in order to
describe the classical behavior at early times after the Planck epoch.
Furthermore, eq.~\eqref{rhoeq} yields that
\begin{equation}
\rho_m=\rho_0(a/a_0)^{-3\gamma_m},
\end{equation}
where $\rho_0$ an $a_0$ are constants, and hence $\rho \rightarrow 0$ at late
times while $\rho \rightarrow \infty$ at early times.

Note that the above qualitative considerations say nothing about how e.g.
\begin{equation}
r = \frac{\rho_\phi}{\rho_m}
\end{equation}
behaves asymptotically, i.e., if the model is fluid or scalar field
dominated, or neither, asymptotically. Nor does the above say anything about
the role of the so-called attractor solution in a global solution space
setting.

This exemplifies that there is a need for a more careful examination, which
is illustrated by some previous heuristic considerations for a scalar field
with a monomial potential by e.g. Turner~\cite{tur83} and
Mukhanov~\cite{muk05} p. 242, which in turn inspired the rigorous work by
Rendall~\cite{ren07}; in addition de la Macorra and Piccinelli introduced a
new heuristic approach to study dynamics at late times for a scalar field
with a monomial potential and a perfect fluid~\cite{macpic00}; rigorous work
in this context was also obtained for the special case $n=1$ by Giambo and
Miritzis~\cite{giamir10}.\footnote{Some further examples of references that
describe minimally coupled scalar field cosmology in dynamical systems
settings are~\cite{col03,beyesc13,fadetal14}, with additional references
therein.} Nevertheless, this still leaves room for improvements and
extensions, and, as will be shown in this paper, it is possible to shed light
on interesting previously neglected physical and mathematical aspects.

The main purpose of this paper in, primarily, mathematical cosmology is
two-fold: Firstly, to obtain a global visual picture of the solutions space,
thus, e.g., situating the so-called attractor solution in a global solution
space context. Secondly, to prove issues concerning asymptotical behavior at
late and early times. This includes introducing averaging techniques to
determine late time behavior, generalizing and simplifying earlier proofs in
the literature, and using center manifold theory to rigorously derive
approximations for the attractor solution at early times, as well as
clarifying the physically important issue of asymptotic self-similarity.

The outline of the paper is as follows. In the next section we introduce our
new three-dimensional dynamical systems reformulation of the field equations
on a relatively compact state space. We also present two other complementary
dynamical systems formulations of the field equations, which allow us to
effectively obtain approximations for the attractor solution. In
Section~\ref{sec:globaldynsys} we apply global and local dynamical systems
techniques to obtain a complete and illustrative picture of the solution
space and its properties, including asymptotics. In particular, we introduce
averaging techniques in our global dynamical systems setting, which allows us
to prove the following theorem:
\begin{theorem}\label{theorem}
\begin{itemize}
\emph{
\item[(i)] If $\gamma_m - \frac{2n}{n+1}>0$, then $r=\rho_\phi/\rho_m
    \rightarrow \infty$ for all solutions with $\rho_\phi\rho_m >0$,
    which implies that the solutions are future \emph{asymptotically
    scalar field dominated}.
\item[(ii)] If $\gamma_m -\frac{2n}{n+1} < 0$, then $r=\rho_\phi/\rho_m
    \rightarrow 0$ for all solutions with $\rho_\phi\rho_m>0$, and thus
    the solutions in this case are future \emph{asymptotically perfect
    fluid dominated}.
\item[(iii)] If $\gamma_m - \frac{2n}{n+1} = 0$, then $r=\rho_\phi/\rho_m
    \rightarrow \mathrm{const}.$ when $\rho_\phi\rho_m >0$, and there is
    thus no future scalar field or perfect fluid dominance.}
\end{itemize}
\end{theorem}
It should be pointed out that similar conclusions have been reached
heuristically with quite different arguments in e.g.~\cite{macpic00}.
Furthermore, Giambo and Miritzis gave a proof for $n=1$ for the cases (i) and
(ii) in~\cite{giamir10} (in the case of general relativity). However, apart
from that our proof rigorously generalizes previous results, %we believe that
our method can, in principle, be modified to treat even more general
situations. Moreover, we tie our results to the global dynamical systems
picture and discuss their physical implications, e.g., situating them in the
context of future manifest asymptotic self-similarity breaking. In
Section~\ref{sec:attrappr} we focus on the attractor solution, where we
introduce and compare several approximation schemes, such as center manifold
and slow-roll based expansions  and Pad{\'e}  approximants, in order to
describe it quantitatively. Finally, Section~\ref{sec:concl} contains some
general remarks, e.g. about the de Sitter solution on the unphysical boundary
of the state space.
%Finally, Appendix~\ref{app:proof} contains {\bf an alternative and simpler
%proof of the future asymptotic behavior, but only for the two cases where
%either the scalar field or the fluid dominates}.

%%%%%%%%%%%%%%%%%%%%%%%%%%%%%%%%%%%%%%%%%%%%%%%%%%%%%%%%%%%%%%%%%%
\section{Dynamical systems formulations}\label{sec:dynsysappr}
%%%%%%%%%%%%%%%%%%%%%%%%%%%%%%%%%%%%%%%%%%%%%%%%%%%%%%%%%%%%%%%%%%

%---------------------------------------------------------------
\subsection{Global dynamical systems}
%---------------------------------------------------------------

Our main global (i.e. compact) dynamical systems formulation is based on the
dependent variables $T, X, \Sigma_\dagger$, which are defined as follows:
\begin{subequations}\label{vardef1}
\begin{align}
(T,X,\Sigma_\dagger) &= \left(\frac{c}{c + H^{1/n}}, \frac{\lambda\phi}{(6nH^2)^{1/2n}},\frac{\dot{\phi}}{\sqrt{6}H}\right),\\
(H,\phi,\dot{\phi}) &= \left(c^n\tilde{T}^{-n},\sqrt{6}\tilde{T}^{-1}X,\sqrt{6}c^n\tilde{T}^{-n}\Sigma_\dagger\right),%\\
%(H,\phi,\dot{\phi},\rho_m) &= \left(\frac{c^n}{\tilde{T}^{n}},\sqrt{6}\frac{X}{\tilde{T}},\sqrt{6}c^n\frac{\Sigma_\dagger}{\tilde{T}^{n}},3c^{2n}\frac{\Omega_m}{\tilde{T}^{2n}}\right),
\end{align}
where
\begin{equation}
\tilde{T} = \frac{T}{1-T}, \qquad
c = \left(\frac{6^{n-1}}{n}\right)^{\frac{1}{2n}}\lambda .
\end{equation}
\end{subequations}
%
%
%\begin{subequations}\label{vardef1}
%\begin{alignat}{2}
%T &= \frac{c}{c + H^{1/n}}, &\qquad H &= c^n\left(\frac{1-T}{T}\right)^n,\\
%X &= \frac{\lambda\phi}{(6nH^2)^{1/2n}}, &\qquad \phi &= \sqrt{6}\left(\frac{1-T}{T}\right)X,\\
%\Sigma_\dagger &= \frac{\dot{\phi}}{\sqrt{6}H}, &\qquad \dot{\phi} &=
%\sqrt{6}c^n\left(\frac{1-T}{T}\right)^n\Sigma_\dagger,\\
%\Omega_m &= \frac{\rho_m}{3H^2}, &\qquad \rho_m &=
%3c^{2n}\left(\frac{1-T}{T}\right)^{2n}\Omega_m,
%\end{alignat}
%
%where
%
%\begin{equation}
%c = \left(\frac{6^{n-1}}{n}\right)^{\frac{1}{2n}}\lambda .
%\end{equation}
%\end{subequations}
%
In addition it is of interest to define
\begin{equation}
\Omega_\phi = \frac{\rho_\phi}{3H^2} = \Sigma_\dagger^2 + X^{2n}, \qquad \Omega_m = \frac{\rho_m}{3H^2}.
\end{equation}
%
%\rho_m = 3c^{2n}\tilde{T}^{-2n}\Omega_m

To introduce a new suitable time variable we note the following: At early
times it is natural to use a Hubble-normalized time variable $\tau$ defined
by $d\tau/dt = H$, due to that the expansion $\theta = 3H$ provides a natural
variable scale when $\theta \rightarrow \infty$ via the Raychaudhuri
equation, as further discussed in e.g.~\cite{ugg13b}, and references therein
(in an inflationary context $\tau$ is often interpreted as the number of
$e$-folds $N$). At late times the square root of the second derivative of the
potential, $\phi^{n-1}$ (for simplicity we here incorporate $\lambda$ into
$\phi$), provides a natural variable (mass) scale. Due to the Gauss
constraint~\eqref{Gauss1}, which relates a scale given in $\phi$ to one given
in $H$ according to $\phi \sim H^{1/n}$, this scale can be expressed in terms
$H$ according to $H^{1-1/n}$ which leads to a dimensionless time variable
$\check{\tau}$ defined by $d\check{\tau}/dt = \mathrm{constant}\cdot
H^{1-1/n}$, where the constant have the same dimension as $H^{1/n}$. To
incorporate these features in a global dynamical systems setting we introduce
a new time variable $\bar{\tau}$ that interpolates between these two regimes,
\begin{equation}\label{bartaudef}
\frac{d\bar{\tau}}{dt} = H(1-T)^{-1},
\end{equation}
where $d\bar{\tau}/dt \rightarrow H,\, c H^{1-1/n}$ when $\bar{\tau}
\rightarrow -\infty$ and $\bar{\tau} \rightarrow  +\infty$, respectively.

The above leads to the following three-dimensional dynamical system for $(T,
\Sigma_\dagger, X)$:\footnote{The variable $\Sigma_\dagger$ has been used
ubiquitously in the scalar field literature (often denoted by $x$), while $X$
was used in~\cite{reyure10} where it was denoted by $y$, however, as far as
we know, the variable $T$ and the independent variable $\bar{\tau}$ are new.
The reason for the name $\Sigma_\dagger$ is that mathematically this variable
plays a role that is reminiscent to that of Hubble-normalized shear, which is
usually denominated by $\Sigma$ in anisotropic cosmology, for a number of
situations (the subscript $\dagger$ follows the notation
in~\cite{uggetal95}). Thus the present nomenclature is designed to pave the
way for eventually situating the present problem in a broader context than
isotropic scalar field cosmology.}
\begin{subequations}\label{dynsys}
\begin{align}
\frac{dT}{d\bar{\tau}} &= \frac{1}{n}T(1-T)^2(1+q),\label{Teq}\\
\frac{d\Sigma_\dagger}{d\bar{\tau}} &= -(1-T)(2-q)\Sigma_\dagger - nTX^{2n-1},\\
\frac{dX}{d\bar{\tau}} &=\frac{1}{n}(1-T)(1+q)X + T\Sigma_\dagger,
\end{align}
\end{subequations}
where the deceleration parameter, $q$, defined via $\dot{H} = - (1+q)H^2$, is
given by
\begin{equation}\label{qexp1}
q = -1 + \frac32\left(\gamma_\phi\Omega_\phi + \gamma_m\Omega_m\right) = - 1 + 3\Sigma_\dagger^2 + \frac32\gamma_m\Omega_m,
\end{equation}
%
%\begin{subequations}\label{qexp1}
%\begin{align}
%1+q &= \sfrac32(\gamma_\phi\Omega_\phi + \gamma_m\Omega_m) = 3(\Sigma_\dagger^2 + \sfrac12\gamma_m\Omega_m),\\
%2-q &= 3(1 - \sfrac12\gamma_\phi\Omega_\phi - \sfrac12\gamma_m\Omega_m) = 3(1 - \Sigma_\dagger^2 - \sfrac12\gamma_m\Omega_m),
%\end{align}
%\end{subequations}
%
where
\begin{equation}\label{Gauss2}
\Omega_\phi = \Sigma_\dagger^2 + X^{2n}, \qquad \Omega_m = 1 - \Sigma_\dagger^2 - X^{2n} = 1 - \Omega_\phi \geq 0,
\end{equation}
where the last equation follows from the Gauss constraint~\eqref{Gauss1},
while the inequality is due to $\rho_m \geq 0$. Above we have also introduced
an effective equation of state parameter $\gamma_\phi$ for the scalar field
which is defined according to
\begin{equation}\label{qscalar}
\gamma_\phi = 1 + \frac{p_\phi}{\rho_\phi} =
1 + \frac{\sfrac12 \dot{\phi}^2 - \frac{1}{2n}(\lambda\phi)^{2n}}{\sfrac12 \dot{\phi}^2 + \frac{1}{2n}(\lambda\phi)^{2n}}
= \frac{\dot{\phi}^2}{\sfrac12 \dot{\phi}^2 + \frac{1}{2n}(\lambda\phi)^{2n}}.
\end{equation}
%
%which leads to $\dot{H} = -\sfrac{1}{2}(\gamma_\phi\rho_\phi +
%\gamma_m\rho_m)$.
From the above relations it follows that $-1 \leq q \leq 2$. In addition it
is of interest to give the following auxiliary evolution equation for
$\Omega_\phi$:
\begin{equation}\label{Omeq}
\frac{d\Omega_\phi}{d\bar{\tau}} = 3(1-T)(\gamma_m - \gamma_\phi)\Omega_\phi\Omega_m,
\end{equation}
where
\begin{equation}
\gamma_\phi\Omega_\phi = 2\Sigma_\dagger^2 .
\end{equation}

The state space ${\bf S}$ associated with eq.~\eqref{dynsys} is given by a
finite (when $n >1$ deformed) cylinder described by the invariant pure
\emph{scalar field boundary subset}, $\Omega_m=0$ (i.e. $\rho_m=0$), and thus
$\Omega_\phi=1$, which we denote by ${\bf S}|_{\Omega_m=0}$, and $0<T<1$.
From now on, we analytically extend ${\bf S}$ to the state space $\bar{\bf
S}$ by including the unphysical invariant submanifold boundaries $T=0$ and
$T=1$. Although these boundaries are unphysical, we stress that it is
essential to include them since they describe the past and future asymptotic
states, respectively, of all physical solutions.

Note that $\Omega_\phi=0$ (i.e. $\Sigma_\dagger = 0 = X$) and hence
$\Omega_m=1$ is an interior invariant subset, ${\bf S}|_{\Omega_\phi=0}$,
which is just a straight line in the center of the cylinder, describing the
flat FLRW perfect fluid model without a scalar field (this solution appears
as a straight line in the present state space due to that it is a
self-similar solution, where $T$ describes the temporal change in the
dimensional variable $H$). Note also that the dynamical system~\eqref{dynsys}
is invariant under the discrete symmetry $(X,\Sigma_\dagger) \rightarrow
-(X,\Sigma_\dagger)$, leading to a double representation of the physical
solutions when $\Omega_m
>0$, which is a consequence of that the potential is invariant when $\phi
\rightarrow - \phi$.\footnote{The system~\eqref{dynsys} is differentiable for
non-integer $n$ when $n>1$, where the differentiability depends on $n$,
describing problems with potentials $V=\frac{1}{2n}(\lambda|\phi|)^{2n}$,
where $X$ is to be replaced with $|X|$ in~\eqref{dynsys}.}

To describe the dynamics on the scalar field boundary $\bar{\bf
S}|_{\Omega_m=0}$, where $\Omega_\phi=1$, it is useful to introduce a
complementary global formulation, which is based on the following
transformation of $\Sigma_\dagger$ and $X$:
\begin{equation}\label{thetadef}
\Sigma_\dagger = F(\theta)\sin\theta, \qquad X = \cos\theta, \qquad
F(\theta) = \sqrt{\frac{1-\cos^{2n}{\theta}}{1-\cos^{2}{\theta}}}.% \in [1,\sqrt{n}],
\end{equation}
This leads to the following regular \emph{unconstrained} two-dimensional
dynamical system:
\begin{subequations}\label{dynsysB}
\begin{align}
\frac{dT}{d\bar{\tau}} &= \frac{3}{n}T(1-T)^2(1-\cos^{2n}{\theta}), \\
\frac{d\theta}{d\bar{\tau}} &= - TF(\theta) - \frac{3}{2n}(1-T)F^2(\theta)\sin 2\theta .
\end{align}
\end{subequations}
In this case the deceleration parameter $q$ is given by
\begin{equation}
q = 2 - 3\cos^{2n}{\theta}.
\end{equation}
The system~\eqref{dynsysB} constitutes a generalization of the system used
in~\cite{alhugg15}. Note that for $n>1$ the present $\theta$ variable is not
the same $\theta$ as that in~\cite{ren07}, which in turn was based on the
variables used in~\cite{beletal85a}.

%---------------------------------------------------------------
\subsection{Complementary non-bounded dynamical systems}\label{sec:early}
%---------------------------------------------------------------

We here introduce two complementary dynamical systems on unbounded state
spaces that are useful for describing the dynamics at early times. The first
system is based on the dependent variables $\tilde{T},X,\Sigma_\dagger$ and
the independent variable $\tau$, where we recall that $\tilde{T}$ and $\tau$
are defined by
\begin{equation}\label{bartaudef}
\tilde{T} = \frac{T}{1-T} =c H^{-1/n}, \qquad \frac{d\tau}{dt} = H,
\end{equation}
where $\tau$ can be viewed as the number of $e$-folds $N$, i.e., $\tau=N$.
This leads to the dynamical system:
\begin{subequations}\label{dynsysearly}
\begin{align}
\frac{d\tilde{T}}{d\tau} &= \frac{1}{n}\tilde{T}(1+q),\label{tildeTeq}\\
\frac{d\Sigma_\dagger}{d\tau} &= -(2-q)\Sigma_\dagger - n\tilde{T}X^{2n-1},\\
\frac{dX}{d\tau} &=\frac{1}{n}(1+q)X + \tilde{T}\Sigma_\dagger,
\end{align}
\end{subequations}
where $q$ is still given by~\eqref{qexp1},~\eqref{Gauss2}.\footnote{In the
special case $\Omega_m=0$ and $n=2$ this system coincides with eq. (16)
in~\cite{kistim08}; incidentally, this model was also the example discussed
by Linde in his paper ``Chaotic inflation''~\cite{lin83}.} It is also useful
to consider auxiliary equations for $\Omega_\phi$ and $r= \rho_\phi/\rho_m =
\Omega_\phi/\Omega_m$:
\begin{subequations}\label{Omeqrtau}
\begin{align}
\frac{d\Omega_\phi}{d\tau} &= 3(\gamma_m - \gamma_\phi)\Omega_\phi\Omega_m, \label{Omegatau}\\
\frac{dr}{d\tau} &= 3(\gamma_m - \gamma_\phi)r.\label{rtau}
\end{align}
\end{subequations}

The second complementary dynamical system concerns the dynamics on the scalar
field boundary $\Omega_m=0$. Expressed in terms of $\tilde{T}$ and $\tau$ the
unconstrained system~\eqref{dynsysB} takes the form:
\begin{subequations}\label{dynsysBearly}
\begin{align}
\frac{d\tilde{T}}{d\tau} &= \frac{3}{n}\tilde{T}(1-\cos^{2n}{\theta}), \\
\frac{d\theta}{d\tau} &= - \tilde{T}F(\theta) - \frac{3}{2n}F^2(\theta)\sin 2\theta .
\end{align}
\end{subequations}
Note that the above systems share the same equations as the previous ones
with bounded state spaces on the invariant boundary subset $\tilde{T}=0=T$.

%%%%%%%%%%%%%%%%%%%%%%%%%%%%%%%%%%%%%%%%%%%%%%%%%%%%%%%%%%%%%%%%%%
\section{Global dynamical systems analysis}\label{sec:globaldynsys}
%%%%%%%%%%%%%%%%%%%%%%%%%%%%%%%%%%%%%%%%%%%%%%%%%%%%%%%%%%%%%%%%%%

It follows from~\eqref{dynsys} that
\begin{equation}\label{mon}
\left. \frac{dT}{d\bar{\tau}}\right|_{1+q= 0} = 0,\qquad
\left. \frac{d^2T}{d\bar{\tau}^2}\right|_{1+q = 0} = 0,\qquad
\left. \frac{d^3T}{d\bar{\tau}^3}\right|_{1+q = 0} = 6nT^3(1-T)^2,
\end{equation}
on ${\bf S}$ (note that since we have assumed that $\gamma_m>0$ it follows
from eq.~\eqref{qexp1} that $q+1=0$ only when $\Omega_m=0$ and
$\gamma_\phi=0$). Due to that $q+1\geq0$, it follows from~\eqref{Teq}
and~\eqref{mon} that $T$ is a monotonically increasing function on ${\bf S}$
(although eq.~\eqref{mon} shows that solutions on the scalar field boundary
have inflection points when $1+q=0$) and hence $T$ can be viewed as a time
variable if one is so inclined. As a consequence all orbits (i.e. solution
trajectories) in ${\bf S}$ originate from the invariant subset boundary
$T=0$, which is associated with the asymptotic (classical) initial state, and
end at the invariant subset boundary $T=1$, which corresponds to the
asymptotic future, and therefore all fixed points are located at $T=0$ and
$T=1$.\footnote{A fixed point, sometimes called an equilibrium, critical, or
stationary point, is a point $x_0$ in the state space of a dynamical system
$\dot{x} = f(x)$ for which $f(x_0)=0$.} It also follows from the monotonicity
of $T$ that the past (future) attractor resides on $T=0$ ($T=1$).

The equations on the subset $T=0$ (or, equivalently $\tilde{T}=0$) are given
by
\begin{subequations}\label{dynsysT0}
\begin{align}
\frac{d\Sigma_\dagger}{d\bar{\tau}} &= -(2-q)\Sigma_\dagger ,\\
\frac{dX}{d\bar{\tau}} &=\frac{1}{n}(1+q)X,
\end{align}
\end{subequations}
as follows from~\eqref{dynsys}, or, equivalently~\eqref{dynsysearly} (with
$\bar{\tau}$ replaced with $\tau$), where $q$ is given by~\eqref{qexp1}
and~\eqref{Gauss2}. It follows that the state space on $T=0$ is divided into
four sectors defined by the invariant subsets $\Sigma_\dagger =0$ and $X=0$.
The intersection of these subsets with $\Omega_m=0$ and with each other yield
five fixed points on $T=0$:
\begin{subequations}\label{dynsysT0}
\begin{alignat}{2}
\mathrm{M}_\pm\!\!: \quad X &=0, &\quad \Sigma_\dagger &= \pm 1, \\
\mathrm{dS}_\pm\!\!:\quad X &=\pm 1, &\quad \Sigma_\dagger &= 0, \\
\mathrm{FL}_0\!\!:\quad X &=0, &\quad \Sigma_\dagger &= 0,
\end{alignat}
\end{subequations}
where $\mathrm{M}_\pm$ are two equivalent fixed points for which $\Omega_m=0$
and $q=2$ (and $\gamma_\phi=2$), i.e., they are associated with a massless
scalar field state, while the two equivalent fixed points $\mathrm{dS}_\pm$,
for which $\Omega_m=0$ and $q=-1$ (and $\gamma_\phi=0$) correspond to a
de~Sitter state.\footnote{Note that the present de Sitter fixed points are
distinct from de Sitter states that are associated with potentials that admit
situations for which $dV/d\phi=0$ for some constant finite value of $\phi$
for which both $V$ and $H$ have constant, bounded, and positive values. In
contrast the present de Sitter states correspond to the limits $\dot{\phi}
=0$, $(\phi,V,H) \rightarrow (\pm\infty,\infty,\infty)$, and therefore reside
on the unphysical boundary $T=0$.} The fixed point $\mathrm{FL}_0$ gives
$\Omega_m=1$ and $q = \sfrac12(3\gamma_m - 2)$ and corresponds to the flat
perfect fluid Friedman model.

\begin{figure}[ht!]
\begin{center}
%\subfigure[$V(\phi)=\frac{1}{2}m^2\phi^2$, $\gamma_m=1$]{\label{fig:Phi2Dust}
%\includegraphics[width=0.4\textwidth]{Phi2_Dust_T0Subset_final.pdf}}
\subfigure[$V(\phi)=\frac{1}{2}m^2\phi^2$, $\gamma_m=\frac{4}{3}$]{\label{fig:Phi2RadiationT0}
\includegraphics[width=0.35\textwidth]{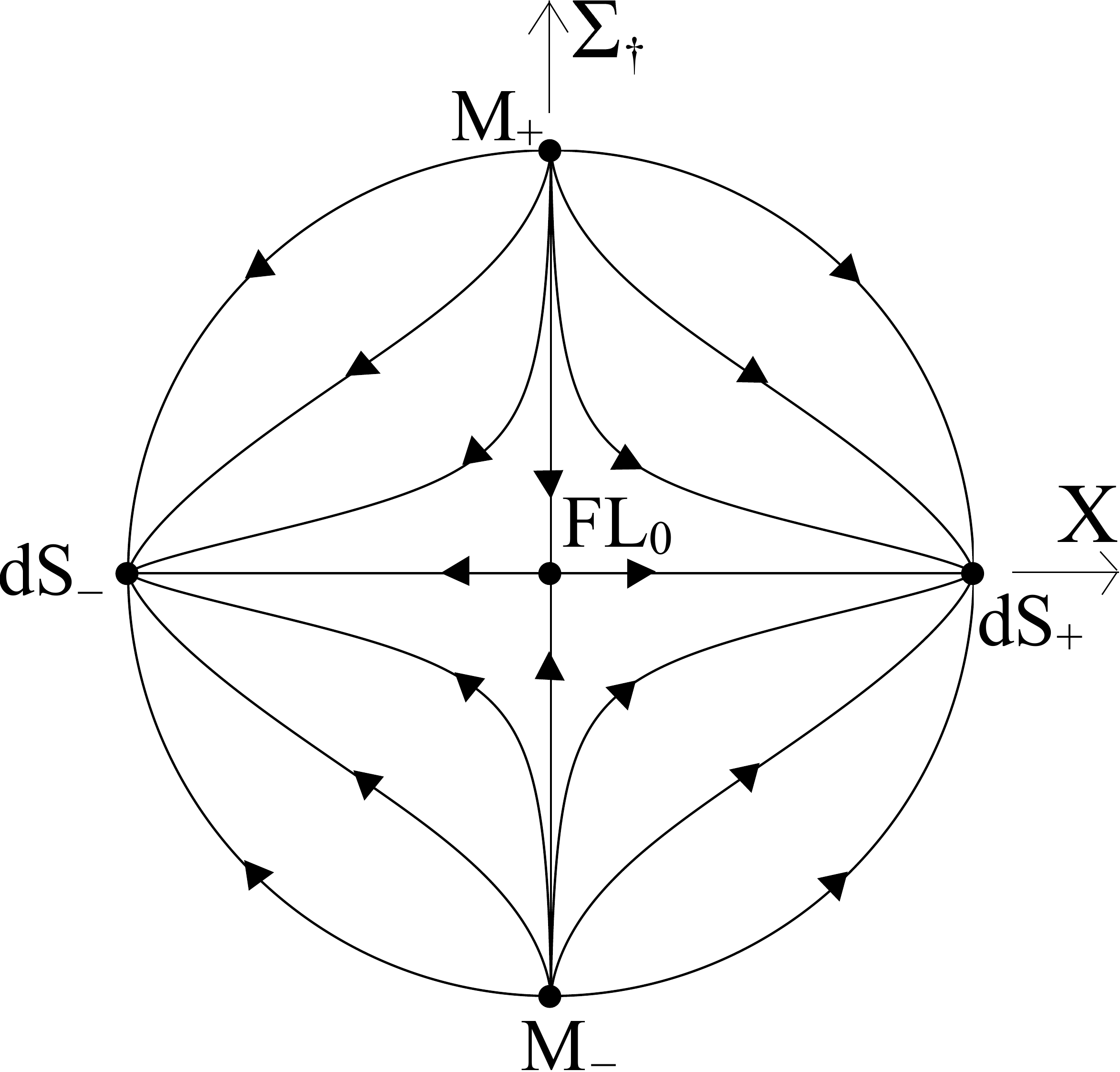}} \qquad
\subfigure[$V(\phi)=\frac{1}{4}(\lambda \phi)^4$, $\gamma_m=1$]{\label{fig:Phi4DustT0}
\includegraphics[width=0.35\textwidth]{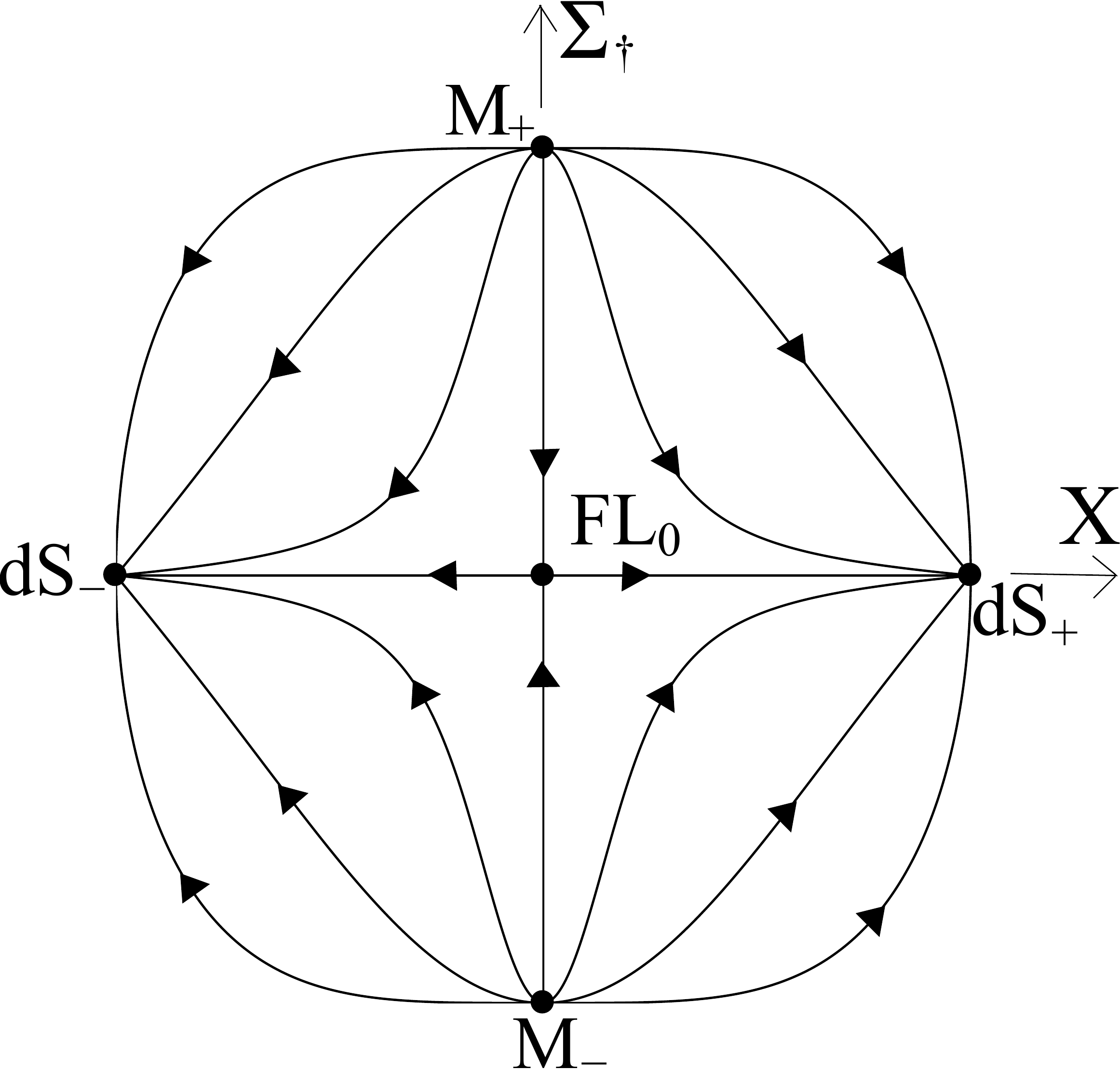}}
%\subfigure[$V(\phi)=\frac{1}{4}(\lambda \phi)^4$, $\gamma_m=\frac{4}{3}$]{\label{fig:Phi2Radiation}
%\includegraphics[width=0.4\textwidth]{Phi4_Radiation_T0Subset_final.pdf}}
\end{center}
\caption{The invariant $T=0$ boundary subset of $\bar{\bf S}$ for monomial potentials and a perfect fluid with
a linear equation of state. The flows are topologically equivalent for all monomial potentials with $n \geq 1$
and perfect fluids with $0<\gamma_m<2$. Depicted are two examples.
%: A scalar field potential
%$V(\phi)=\frac{1}{2}m^2\phi^2$ and
%$V(\phi)=\frac{1}{4}(\lambda \phi)^4$ and a perfect fluid with an equations of state
%$\gamma_m=1$ and $\frac43$, respectively.
}\label{fig:T0}
\end{figure}

As shown below, the two fixed points $\mathrm{M}_\pm$ are sources on
$\bar{\bf S}$; the fixed points $\mathrm{dS}_\pm$ are sinks \emph{on} $T=0$,
but they also have one zero eigenvalue that corresponds to a one-dimensional
unstable center submanifold on ${\bf S}|_{\Omega_m=0}$, i.e., one solution,
called an attractor solution, originates from each fixed point
$\mathrm{dS}_\pm$ into ${\bf S}$ on the pure scalar field boundary subset
$\Omega_m=0$. The fixed point $\mathrm{FL}_0$ is a saddle that gives rise to
a 1-parameter set of solutions entering ${\bf S}$ (the associated unstable
tangent space is given by $\Sigma_\dagger =0$), one being the perfect fluid
solution given by $\Omega_\phi=0, \Omega_m=1$. The system~\eqref{dynsysT0}
admits the following conserved quantity when $\Omega_m = 1 - \Sigma_\dagger^2
- X^{2n}>0$:
\begin{equation}
\Sigma_\dagger^{\gamma_m}\,X^{(2 - \gamma_m)n}\,\Omega_m^{-1} = \mathrm{const}.,
\end{equation}
which determines the solution trajectories on the $T=0$ subset, see
Figure~\ref{fig:T0}.

\begin{figure}[ht!]
\begin{center}
\subfigure[$V(\phi)=\frac{1}{2}m^2\phi^2$]{\label{fig:Phi2T1}
\includegraphics[width=0.35\textwidth]{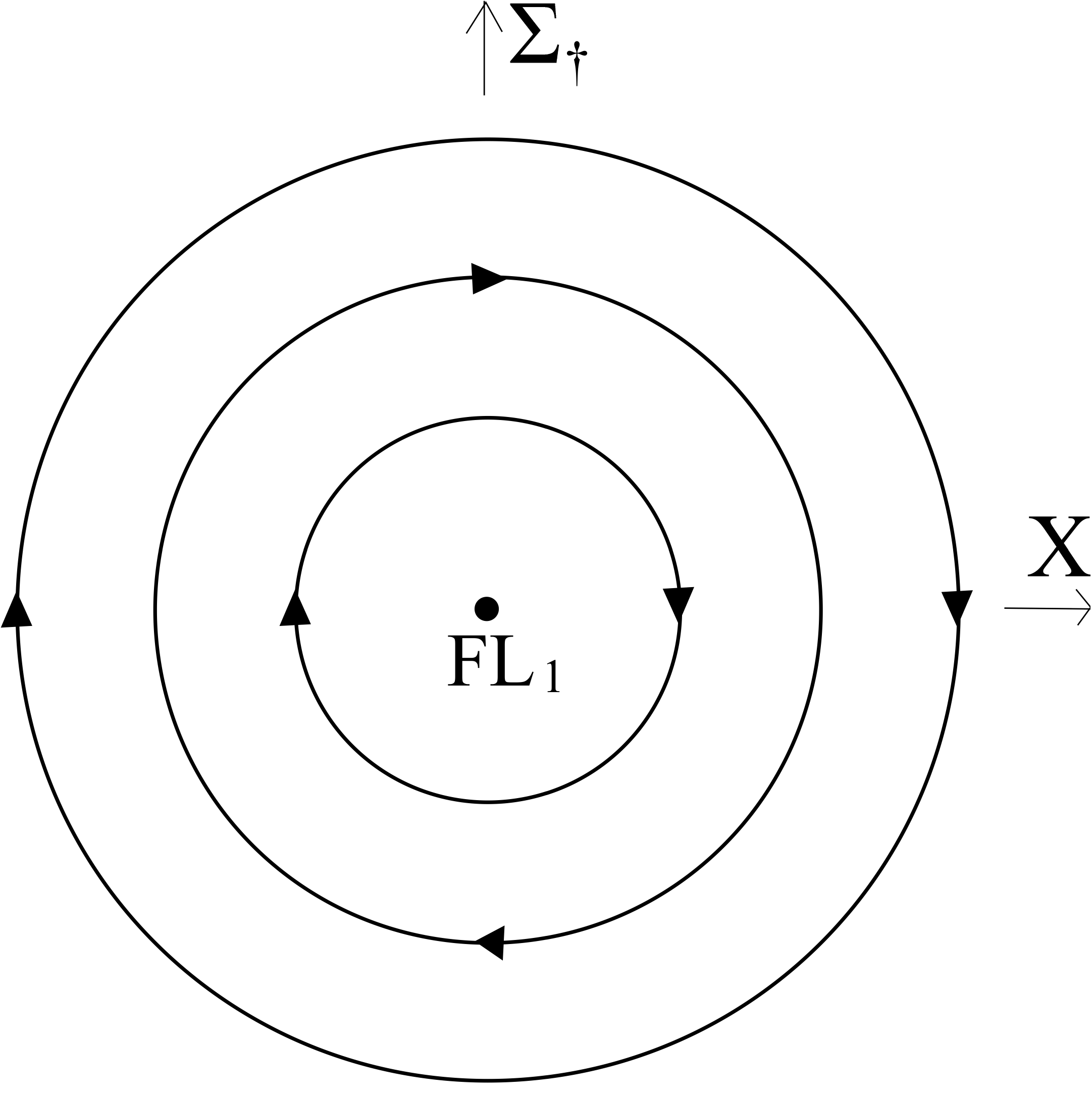}}
\subfigure[$V(\phi)=\frac{1}{4}(\lambda\phi)^4$]{\label{fig:Phi4T1}
\includegraphics[width=0.35\textwidth]{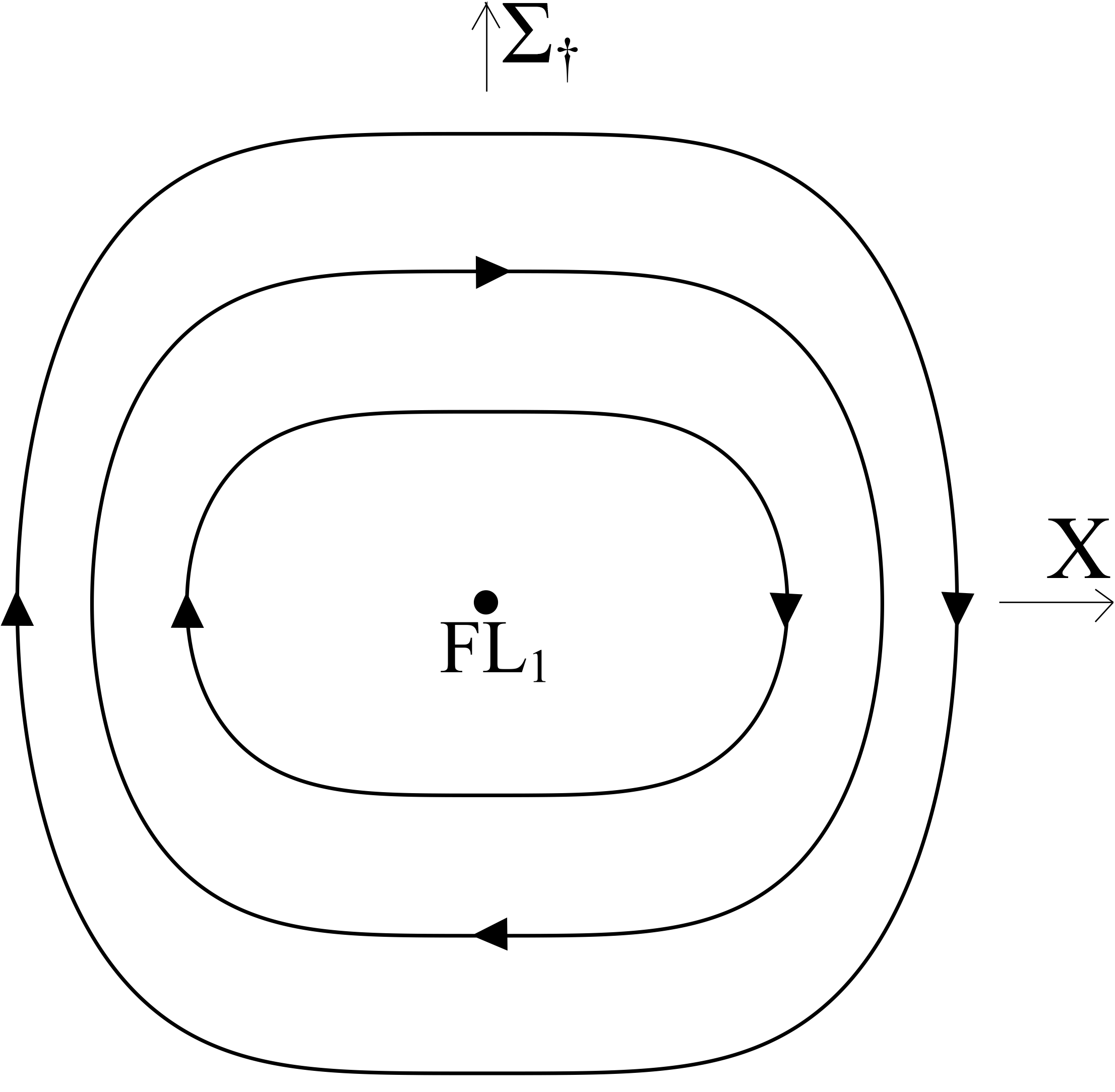}}
\end{center}
\caption{The invariant $T=1$ boundary subset of $\bar{\bf S}$ for two examples:
The potentials $V(\phi)=\frac{1}{2}m^2\phi^2$ and $V(\phi)=\frac{1}{4}(\lambda\phi)^4$
and a perfect fluid (with arbitrary equation of state $0<\gamma_m<2$).}\label{fig:T1}
\end{figure}

The equations on the subset $T=1$ are given by
\begin{subequations}\label{dynsysT1}
\begin{align}
\frac{d\Sigma_\dagger}{d\bar{\tau}} &= -nX^{2n-1},\\
\frac{dX}{d\bar{\tau}} &= \Sigma_\dagger.
\end{align}
\end{subequations}
This system has a non-hyperbolic fixed point,\footnote{A fixed point is
hyperbolic if the linearization of the dynamical system at the fixed point is
a matrix that possesses eigenvalues with non-vanishing real parts; if the
linearization leads to one or more eigenvalues with vanishing real parts it
is said to be non-hyperbolic.}
\begin{equation}
\mathrm{FL}_1\!\!:\quad X =0, \quad \Sigma_\dagger = 0,
\end{equation}
with three zero eigenvalues. Fortunately $\mathrm{FL}_1$ resides at the
intersection of two invariant subsets: the invariant subset $T=1$ and the
invariant perfect fluid subset $\Omega_\phi=0, \Omega_m=1$, and thus, since
$T$ is monotone in ${\bf S}$, $\mathrm{FL}_1$ attracts at least this orbit.
\emph{On} $T=1$ $\mathrm{FL}_1$ is conveniently analyzed by considering
eq.~\eqref{Omeq} on $T=1$, which yields that
\begin{equation}
\Omega_\phi = \Sigma_\dagger^2 + X^{2n} = \mathrm{const},
\end{equation}
which is an integral of~\eqref{dynsysT1}, i.e., the subset $T=1$ is foliated
with periodic orbits surrounding the fixed point $\mathrm{FL}_1$, see
Figure~\ref{fig:T1}.

To make further progress as regards the global properties of the solution
space we need to consider the asymptotic dynamics at early and late times.

%---------------------------------------------------------------
\subsection{Asymptotic dynamics at early times}\label{sec:early}
%---------------------------------------------------------------

The linearization of the system~\eqref{dynsys} at the fixed points
$\mathrm{FL}_0$ and $\mathrm{M}_\pm$ is conveniently described as follows:
\begin{subequations}
\begin{xalignat}{2}
\frac{1}{T}\left. \frac{dT}{d\bar{\tau}}\right|_{\mathrm{FL}_0} &= \frac{3}{2n}\gamma_m, & \quad
\frac{1}{T}\left. \frac{dT}{d\bar{\tau}}\right|_{\mathrm{M}_\pm} &= \frac{3}{n},\\
\frac{1}{\Sigma_\dagger}\left.
\frac{d\Sigma_\dagger}{d\bar{\tau}}\right|_{\mathrm{FL}_0} &=
-\frac{3}{2}\left(2 - \gamma_m\right), &\quad
\frac{1}{1 \pm \Sigma_\dagger}\left. \frac{d(1 \pm \Sigma_\dagger)}{d\bar{\tau}}\right|_{\mathrm{M}_\mp} &= 3(2 - \gamma_m),\\
\frac{1}{X}\left. \frac{dX}{d\bar{\tau}}\right|_{\mathrm{FL}_0} &=
\frac{3}{2n}\gamma_m, &\quad \frac{1}{X}\left.
\frac{dX}{d\bar{\tau}}\right|_{\mathrm{M}_\pm} &= \frac{3}{n},
\end{xalignat}
\end{subequations}
where the right hand sides constitute the eigenvalues of the fixed points.
Hence $\mathrm{FL}_0$ is a hyperbolic saddle, with an unstable manifold
tangential to $\Sigma_\dagger =0$, i.e., there is a 1-parameter set of
solutions that originate from $\mathrm{FL}_0$ entering the state space ${\bf
S}$ tangentially to $\Sigma_\dagger =0$. The fixed points $\mathrm{M}_\pm$
are hyperbolic sources; it follows from the invariant submanifold structures
that there exists a 2-parameter set of solutions entering the interior of the
cylinder with $\Omega_m>0$ from each fixed point $\mathrm{M}_\pm$, while a
1-parameter set of solutions originate from each fixed point $\mathrm{M}_\pm$
into the boundary subset ${\bf S}|_{\Omega_m=0}$.\footnote{The above results
follow from the Hartman-Grobman theorem, which states that in a neighborhood
of a hyperbolic fixed point the full nonlinear dynamical system and the
linearized system are topologically equivalent, see
e.g.~\cite{cra91,alhugg15}.}

\emph{On} the subset $T=0$ linearization of the system~\eqref{dynsysT0} gives
\begin{subequations}
\begin{align}
\frac{1}{\Sigma_\dagger}\left. \frac{d\Sigma_\dagger}{d\bar{\tau}}\right|_{\mathrm{dS}_\pm} &= -3,\\
\frac{1}{1 \mp X}\left. \frac{d(1 \mp X)}{d\bar{\tau}}\right|_{\mathrm{dS}_\pm} &= -3\gamma_m,
\end{align}
\end{subequations}
and hence ${\mathrm{dS}_\pm}$ are hyperbolic sinks on $T=0$, as illustrated
in Figure~\ref{fig:T0}. In the full state space, however, each equivalent
fixed point ${\mathrm{dS}_\pm}$ have an additional zero eigenvalue associated
with a one-dimensional so-called center manifold. Fortunately, the center
direction lies on the ${\bf S}|_{\Omega_m=0}$ subset, and hence we can
investigate the center manifold by means of the unconstrained
system~\eqref{dynsysB}, which will be
done in Section~\ref{sec:attrappr}. %Without loss of generality we can restrict our
%analysis to the fixed point $\mathrm{dS}_+$ at $\theta=0$ on $\bar{\bf
%S}|_{\Omega_m=0}$ (due to the discrete symmetry the results for
%$\mathrm{dS}_-$ are completely analogous to those of $\mathrm{dS}_+$). A
%linear analysis of $\mathrm{dS}_+$ in the $(T,\theta)$ system~\eqref{dynsysB}
%yields
%
%\begin{subequations}\label{linearanalysis}
%\begin{align}
%E^s &= \{(T,\theta)|T=0\},\\
%E^c &= \{(T,\theta)|\sqrt{n}T + 3\theta =0\},
%\end{align}
%\end{subequations}
%
%where we follow the nomenclature in~\cite{cra91} and use $E^s$ and $E^c$ to
%denote the tangential stable and center subspaces, respectively.
There we show that the center manifold of each (equivalent) fixed point
${\mathrm{dS}_\pm}$ corresponds to a single solution that enters the state
space ${\bf S}|_{\Omega_m=0}$ (we will even obtain approximate expressions
for this solution), and this solution, which hence resides on ${\bf
S}|_{\Omega_m=0}$, is what is often referred to as the `attractor' solution.
In the full state space $\bar{\bf S}$ the fixed points $\mathrm{dS}_\pm$ are
thus center-saddles.

From these considerations, in combination with the monotonicity of $T$, it
follows that all solutions are \emph{past asymptotically self-similar} in the
sense that all physical geometrical scale-invariant observables, such as the
deceleration parameter $q$, are asymptotically constant. However, there is a
twist to this. The geometry of a flat FLRW pure perfect fluid cosmology with
a linear equation of state and the geometry of a pure massless scalar field
are geometries that admit a proper spacetime transitive homothety group, and
such spacetimes are invariant under scalings of the spacetime coordinates.
This is the underlying reason why they can be represented by fixed points,
but not all fixed points are associated with geometries admitting proper
spacetime transitive homothety groups, as exemplified by the de Sitter
spacetime. For these spacetimes homothetic scale-invariance is broken by the
dimensional cosmological constant, but the 1-parameter set of de Sitter
spacetimes (parameterized by $\Lambda$) admits a scaling (self-similar)
property that scales $\Lambda$ (i.e., a scaling that maps one de Sitter
spacetime to another with a different $\Lambda$), and it is due to this
scaling property de Sitter spacetimes can appear as fixed points.

In the present case the fixed points are not in the interior physical state
space, but on the unphysical boundary, but they are nevertheless
characterized by e.g. the same value of $q$ as the associated physical
spacetime. The fact that they in the present context appear on the unphysical
boundary has consequences, which we will come back to in a discussion about
the de Sitter fixed points in Section~\ref{sec:concl}. Finally we point out
that, apart from a set of measure zero, all solutions originate from a
massless scalar field state and hence the present models are \emph{past
generically massless scalar field dominated}.

%---------------------------------------------------------------
\subsection{Asymptotic dynamics at late times}\label{sec:late}
%---------------------------------------------------------------

The equations on the $T=1$ subset, i.e. equation~\eqref{dynsysT1}, are equivalent to
that of~\eqref{KG} when setting $H=0$, i.e. this problem is \emph{exactly}
that of an anharmonic oscillator (when $n>1$; for $n=1$ the problem is that
of a harmonic oscillator). This can be seen from eq.~\eqref{dynsysT1}, which
yields
\begin{equation}
\frac{d^2X}{d\bar{\tau}^2} + nX^{2n-1} = 0.
\end{equation}
We will now apply the approximate ideas in Mukhanov~\cite{muk05} to the
present exact problem of an anharmonic oscillator. We therefore first
multiply the above equation with $X$ and rewrite it as
\begin{equation}\label{Xaeq}
\frac{d}{d\bar{\tau}}\left(X\frac{dX}{d\bar{\tau}}\right) - \left(\frac{dX}{d\bar{\tau}}\right)^2 + nX^{2n} = 0.
\end{equation}
Each periodic orbit is characterized by a constant value of $\Omega_\phi$ and
has an associated time period $P = P(\Omega_\phi)$. The time average of a
function $f$ over a period for a periodic orbit characterized by
$\Omega_\phi$ is given by
\begin{equation} \label{defaverage}
\left<f\right>_{\Omega_\phi} =
\int_{\bar{\tau}_0}^{\bar{\tau}_0 + P(\Omega_\phi)} f
d\bar{\tau}/P(\Omega_\phi).
\end{equation}
%Applying the average to several periods and then
%let the number of periods go to infinity leads to
%$\left<f\right>_{\Omega_\phi} = \lim_{\bar{\tau} \rightarrow \infty}
%\left(\bar{\tau}^{-1} \int_{\bar{\tau}_0}^{\bar{\tau}} f
%d\bar{\tau}'\right)$.
Taking the time average of eq.~\eqref{Xaeq} for a
periodic orbit gives
\begin{equation}\label{avrel}
\left<\left(\frac{dX}{d\bar{\tau}}\right)^2\right> = \left<\Sigma_\dagger^2\right> = n\left<X^{2n}\right>,
\end{equation}
where we for notational convenience from now on drop the subscript
$\Omega_\phi$. Again, note that in contrast to the result in~\cite{muk05},
this is an exact relation \emph{on} the subset $T=1$. Using this result on
$T=1$ for a periodic orbit in the definition of $\gamma_\phi$ yields
\begin{equation}
\left<\gamma_\phi\right> = \left<\frac{2\Sigma_\dagger^2}{\Omega_\phi}\right> =
\frac{2\left<\Sigma_\dagger^2\right>}{\Omega_\phi} = \frac{2\left<\Sigma_\dagger^2\right>}{\left<\Sigma_\dagger^2 + X^{2n}\right>}
= \frac{2\left<\Sigma_\dagger^2\right>}{\left<\Sigma_\dagger^2\right> + \left<X^{2n}\right>},
\end{equation}
which together with~\eqref{avrel} leads to
\begin{equation}\label{gammaav}
\left<\gamma_\phi\right> = \frac{2n}{n+1}
\end{equation}
on the subset $T=1$, i.e., $\left<\gamma_\phi\right>_{\Omega_\phi} =
\left<\gamma_\phi\right>$ is independent of $\Omega_\phi$. It therefore
follows that on average, in the above sense, e.g. $n=1$ and $n=2$ on $T=1$
correspond to dust and radiation, respectively. Note that the
result~\eqref{gammaav} coincides with the approximate heuristic results using
proper time given in~\cite{tur83} and~\cite{muk05}; see also~\cite{ren07} for
a quite different precise definition of $\left<\gamma_\phi\right>$, which
still yields~\eqref{gammaav}.

Before continuing it is instructive to consider a model that consists of two
perfect fluids with constant equation of state parameters $\gamma_1$ and
$\gamma_2$. Then $r_m=\rho_1/\rho_2 \propto a^{3(\gamma_2 - \gamma_1)}$ (this
expression follows from that $\rho \propto a^{-3\gamma}$, but it can also be
obtained from the equation $dr_m/d\tau = 3(\gamma_2-\gamma_1)r_m$). Since it
is not difficult to show that $a\rightarrow \infty$ toward the future it
follows that $r_m\rightarrow \infty$ if $\gamma_2>\gamma_1$; $r_m\rightarrow
0$ if $\gamma_2<\gamma_1$; $r_m\rightarrow \mathrm{const}.$ if
$\gamma_2=\gamma_1$, i.e., the fluid with the softest equation of state
dominates toward the future.

\emph{Assuming} that asymptotically $\gamma_\phi$ can be replaced with the
asymptotic averaged result $\left<\gamma_\phi\right> = \frac{2n}{n+1}$ in
eq.~\eqref{rtau} results in
\begin{equation}
\frac{dr}{d\tau} = 3\left(\gamma_m - \left<\gamma_\phi\right>\right)r = 3\left(\gamma_m - \frac{2n}{n+1}\right)r,
\end{equation}
which suggests Theorem~\ref{theorem}, given in the Introduction, but this is
of course no proof. Next we introduce averaging techniques that are
subsequently used to provide the proof of this theorem.

\subsubsection*{Averaging}

Standard averaging techniques and theorems can be found in Chapter 4
in~\cite{GuckHo} (the periodic case) and in~\cite{SanVerMu} (the general
case). In standard averaging theory, a perturbation parameter $\varepsilon$
plays the key role: roughly speaking, a differential equation of the form
$\dot{x}=\varepsilon f(x,t,\varepsilon)$ for $\varepsilon>0$ is approximated
by the averaged equation at $\varepsilon=0$, i.e., $\dot{\bar{y}}=\varepsilon
\la f(\bar{y},.,0)\ra $, where the average $\la.\ra$ is defined in
eq.~\eqref{defaverage}. Furthermore, the error $\vert x-\bar{y}\vert$ has to
be controlled. In the problem at hand, we consider the differential equations
in the variables $\Omega_{\phi}$ and $T$, where the role of the parameter
$\varepsilon$ is played by $1-T$. Therefore, after setting
\begin{equation}
\varepsilon=1-T,
\end{equation}
we have to prove an averaging theorem for the case where $\varepsilon$ is not
a parameter, but a variable that slowly goes to zero. The evolution equation
of $\Omega_{\phi}$ is given in eq.~\eqref{Omeqrtau}, which in terms of
$\varepsilon$ takes the form
\begin{equation}
\frac{d\Omega_{\phi}}{d \bar{\tau}} = 3 \varepsilon (\gamma_m -
\gamma_{\phi})\Omega_{\phi} (1-\Omega_{\phi}),
\end{equation}
where $\gamma_{\phi}=2 \Sigma_\dagger ^2/\Omega_\phi$. This formulation is
problematic due to that $\gamma_{\phi}$ is not well-defined when
$\Omega_\phi$ is zero. We therefore use the following formulation:
\begin{subequations}\label{Ophieps}
\begin{align}
\frac{d\Omega_{\phi}}{d \bar{\tau}} &=
3 \varepsilon (\gamma_m \Omega_\phi - 2\Sigma_\dagger^2) (1-\Omega_{\phi}), \label{sysforaveragingO} \\
\frac{d \varepsilon}{d \bar{\tau}}&= -\frac1n \varepsilon^2 (1- \varepsilon) (1+q), \label{sysforaveraginge}
\end{align}
\end{subequations}
where $(X,\Sigma_\dagger)$ solves the system~\eqref{dynsys},
$\Omega_\phi=\Sigma_\dagger^2+X^{2n}$, and
\begin{equation}
q+1=\frac{3}{2}\left(2 \Sigma_\dagger^2 +
\gamma_m(1-\Omega_\phi)\right).
\end{equation}
The general idea of averaging is to express $\Omega_{\phi}$ as
\begin{equation}\label{varyw}
\Omega_{\phi}=y + \varepsilon w(y, \varepsilon, \bar{\tau}),
\end{equation}
and prove that the evolution of the variable $y$ is  approximated  at first
order by the solution $\bar{y}$ of the averaged equation. For that, consider
the average as defined in eq.~\eqref{defaverage} of the right hand side of
eq.~\eqref{sysforaveragingO}. More precisely, considering an equation of the
form $y'=\varepsilon f(y, \varepsilon, \bar{\tau})
+\mathcal{O}(\varepsilon^2)$ with $\bar{\tau}$-periodic $f(y, 0, \bar{\tau})$
of period $P=P(y)$, the averaged equation is given by $\bar{y}'=  \la f \ra
(y)$, where $\la f \ra (y):= \frac1P \int_0^P f(y, 0,
\bar{\tau})d\bar{\tau}$. According to~\eqref{gammaav}, we have $2\la
\Sigma_\dagger^2 \ra =\la \gamma_\phi\ra \Omega_\phi$, where $\la \gamma_\phi
\ra=\frac{2n}{n+1}$ is a constant that does not depend on $\Omega_\phi$.
Hence the averaged equation reads
\begin{equation} \label{averbary}
\frac{d \bar{y}}{d \bar{\tau}} = 3 \varepsilon (\gamma_m - \la \gamma_\phi \ra) \bar{y} (1- \bar{y}),
\end{equation}
while $w$ in eq.~\eqref{varyw} will be chosen appropriately in the proof.

\subsubsection*{Proof of Theorem~\ref{theorem}}

To prove Theorem~\ref{theorem} we first re-express the theorem in terms of
$\Omega_\phi$:
\begin{itemize}
\item[(i)] If $\gamma_m -\frac{2n}{n+1}>0$, initial conditions with
    positive $\Omega_\phi\ \leq 1$ and $\varepsilon$ converge for
    $\bar{\tau}\rightarrow \infty$ to the outer periodic orbit with
    $\Omega_\phi=1$ tangentially to the slice $\{\varepsilon=0\}$.
\item[(ii)] If $\gamma_m -\frac{2n}{n+1}<0$, initial conditions with
    positive $1-\Omega_\phi$ and $\varepsilon$ converge for
    $\bar{\tau}\rightarrow \infty$ to the center with $\Omega_\phi=0$
    tangentially to the slice $\{\varepsilon=0\}$.
\item[(iii)] If $\gamma_m -\frac{2n}{n+1}=0$, each periodic orbit on the
    slice $\{ \varepsilon=0\}$ attracts a 1-parameter set of trajectories
    with positive initial $\varepsilon$.
\end{itemize}
%
%[Result expressed in terms of $\Omega_\phi$]

\begin{proof}
Let us first derive a differential equation for $y$ by taking the time
derivative of eq.~\eqref{varyw}:
\begin{equation}
\begin{split}
\frac{d\Omega_{\phi}}{d \bar{\tau}} &= \frac{d y}{d \bar{\tau}} + \frac{d\varepsilon}{d \bar{\tau}} w + \varepsilon \left(
\frac{\partial w}{\partial y} \frac{dy}{d\bar{\tau}} + \frac{\partial w}{\partial \bar{\tau}} + \frac{\partial w}{\partial \varepsilon}
\frac{d\varepsilon}{d\bar{\tau}} \right) \\
&= \left(  1+ \varepsilon \frac{\partial w}{\partial y}  \right)
\frac{d y}{d \bar{\tau}} +  \varepsilon \frac{\partial w}{\partial \bar{\tau}} +
\frac{d \varepsilon}{d \bar{\tau}}w + \varepsilon \frac{\partial w }{\partial \varepsilon}  \frac{d \varepsilon}{d \bar{\tau}} .
\end{split}
\end{equation}
On the other hand,
\begin{equation}
\begin{split}
\frac{d\Omega_{\phi}}{d \bar{\tau}} &= 3\varepsilon \left( \gamma_m - \la \gamma_\phi\ra  +
\la \gamma_\phi\ra - \gamma_\phi \right) (y+\varepsilon w)(1-y- \varepsilon w )\\
&= 3\varepsilon  (\gamma_m - \la \gamma_\phi \ra ) y(1-y) +   3  ( \la \gamma_\phi\ra y - 2\Sigma_\dagger^2  )(1-y) \\
& \quad + 3 \varepsilon^2  (\gamma_m - \gamma_\phi) w(1-2y) - 3 \varepsilon^3 (\gamma_m - \gamma_\phi)w^2.
\end{split}
\end{equation}
%

%Given $\Omega_\phi \in [0,1]$, there is a unique periodic orbit $(\Sigma_{\dagger}(\bar{\tau}), X(\bar{\tau}))$ at the top of the cylinder (i.e. at $\varepsilon=0$ or $T=1$) with $\Sigma_\dagger^2+ X^{2n}\equiv \Omega$. Using the periodic function $\Sigma_\dagger (\bar{\tau})$,  we define a  function $\tilde{\gamma}_{\phi}$  by
%\begin{equation}
%\tilde{\gamma_\phi (\Omega_\phi, \bar{\tau})}= \begin{cases}
%\frac{2\Sigma_{\dagger}^2}{\Omega_\phi} & \text{, if } \Omega_\phi\neq 0,\\
%\la \gamma_\phi \ra & \text{, if } \Omega_\phi=0.
%\end{cases}
%\end{equation}
%Note that $\tilde{\gamma_\phi }$ is periodic in $\bar{\tau}$, and $\gamma_\phi - \tilde{\gamma}_\phi= \mathcal{O}(\varepsilon)$ for initial conditions with $\Omega>0$, along whose trajectories the function $\gamma_\phi$ is well-defined.
Let us now set
\begin{equation} \label{defw}
\frac{\partial w}{\partial \bar{\tau}} =  3  ( \la \gamma_\phi\ra y - 2\Sigma_\dagger^2  )(1-y).
\end{equation}
Note that for large times $2\Sigma_\dagger^2$ is well approximated by
periodic functions with an average $\la \gamma_\phi\ra y$. The right hand
side of~\eqref{defw} is for large times almost periodic and has an average
that is zero so that the variable $w$ is bounded.

As a consequence, the differential equation for the variable $y$ takes the
form
\begin{equation}
\begin{split}
\frac{dy}{d\bar{\tau}} =&  \left(1 + \varepsilon \frac{\partial w}{\partial y}  \right)^{-1} \\
& \Bigg\{  3\varepsilon  (\gamma_m - \la \gamma_\phi\ra ) y(1-y)  \\
& + \varepsilon^2  w\left(  3(\gamma_m - \gamma_\phi )(1-2y)  +\frac1n (1-\varepsilon) (1+q)  \right) \\
&  - \varepsilon \frac{\partial w}{\partial \varepsilon} \frac{d \varepsilon}{d \bar{\tau} } -3\varepsilon^3 (\gamma_m - \gamma_\phi)w^2 \Bigg\}.
\end{split}
\end{equation}
Using the fact that $\varepsilon  \frac{\partial w}{\partial \varepsilon}
\frac{d \varepsilon}{d \bar{\tau} }= \mathcal{O}(\varepsilon^3)$ and that
$\left( 1+ \varepsilon \frac{\partial w}{\partial y}  \right)^{-1} = 1 -
\varepsilon \frac{\partial w}{\partial y} + \mathcal{O}(\varepsilon^2)$
results in the following:

\begin{equation} \label{eqfory}
\begin{split}
\frac{d y}{d \bar{\tau}} &=
3\varepsilon (\gamma_m - \la \gamma_\phi\ra ) y(1-y)  \\
& \quad + \varepsilon^2\left\{ 3w(\gamma_m - \gamma_\phi)(1-2y) + \frac1n (1-\varepsilon) (1+q)
- 3\frac{\partial w}{\partial y}(\gamma_m - \la \gamma_\phi \ra)y(1-y) \right\}\\
&\quad + \mathcal{O}(\varepsilon^3).
%w (3(\gamma_m - \gamma_\phi (1-2y) ) +\frac1n (1-\varepsilon) (1+q)) - 3D_yw   (\gamma_m - < \gamma_\phi > ) y(1-y)
%\right)
\end{split}
\end{equation}

Next we have to prove that the solution $y$ of this equation and the solution
$\bar{y}$ of the averaged equation~\eqref{averbary} have the same asymptotics
when $\bar{\tau}\rightarrow +\infty$.
%[explain why the asymptotics of the averaged equation are 0 or 1 depending on the sign of $\gamma_m - \la \gamma_\phi \ra $]
Since the averaged equation~\eqref{averbary} is expected to govern the
dynamics, we first study the late time behavior of the system
\begin{subequations}\label{sysaverage}
\begin{align}
\frac{d \bar{y}}{d \bar{\tau}} &= 3 \varepsilon (\gamma_m - \la \gamma_\phi \ra) \bar{y} (1- \bar{y}),\label{sysaverbary}\\
\frac{d \varepsilon}{d \bar{\tau}}&= -\frac1n \varepsilon^2 (1- \varepsilon) (1+q).\label{sysavereps}
\end{align}
\end{subequations}
After Euler multiplication by $1/\varepsilon$ (or equivalently, a singular change of
time variable $\varepsilon d/d\tau= d/d\bar{\tau}$), this system reads
\begin{subequations}
\begin{align}
\frac{d\bar{y}}{d\tau} &= 3  (\gamma_m - \la \gamma_\phi \ra) \bar{y} (1- \bar{y}), \label{eulerav}\\
\frac{d\varepsilon}{d\tau} &= -\frac1n \varepsilon (1- \varepsilon) (1+q).\label{eulereps}
\end{align}
\end{subequations}

In cases (i) and (ii), for which $\gamma_m - \la \gamma_\phi \ra \neq 0$, the
two fixed points of this system are located at $(\varepsilon=0, \bar{y}=0)$
and $(\varepsilon=0, \bar{y}=1) $. The line $\varepsilon=0 $ is a
heteroclinic orbit between these two fixed points, whose direction depends on
the sign of $\gamma_m - \la \gamma_\phi \ra$.

Undoing the Euler multiplication does not affect the trajectories with
positive $\varepsilon$. On the other hand, in the original averaged
system~\eqref{sysaverage}, the line $\{ \varepsilon=0\}$ is a line of fixed
points. Solutions with positive initial $\varepsilon$ will approach the line
of fixed points at $\varepsilon=0$, but slowly slide along this line as
$\bar{\tau}\rightarrow \infty$ in the direction prescribed by the sign of
$\gamma_m - \la \gamma_\phi \ra $, and go to the left or right fixed point
accordingly.

Next we prove that the variables $y$ and $\Omega_\phi$ follow this evolution.
First note that the sequences $\{\varepsilon_n\}_{n\in \mathbb{N}}$,
$\{\bar{\tau}_n\}_{n\in \mathbb{N}}$, defined as follows,
\begin{equation} \label{bootstrap}
\begin{array}{ll}
\begin{cases}
\bar{\tau}_0=0,\\
\varepsilon_0>0,
\end{cases} &     \qquad
\begin{cases}
\bar{\tau}_{n+1} = \bar{\tau}_n +1/\varepsilon_n,\\
\varepsilon_{n+1}=\varepsilon(\bar{\tau}_{n+1}),
\end{cases}
\end{array}
\end{equation}
have limits
\begin{equation}
\begin{cases}
\lim_{n\to \infty} \bar{\tau}_n = +\infty ,\\
\lim_{n\to \infty} \varepsilon_n= 0 ,
\end{cases}
\end{equation}
since $\varepsilon(\bar{\tau})$ goes to zero when $\bar{\tau}$ goes to
infinity. For a sufficiently small $\varepsilon>0$, eq.~\eqref{eqfory}
guaranties that $y$ is monotone (in- or decreasing, depending on the sign of
the quantity $\gamma_m - \la \gamma_\phi \ra$) and bounded. Hence
$y(\bar{\tau})$ must have a limit when $\bar{\tau}\to \infty$.  Next we
estimate $\zeta(\bar{\tau}):= y(\bar{\tau})-\bar{y}(\bar{\tau})$, where $y$
and $\bar{y}$ are trajectories with identical `initial' conditions at time
$\tau_n$:
\begin{equation}
\begin{split}
\vert \zeta(\bar{\tau})\vert &=  \big| \int_{\bar{\tau}_n}^{\bar{\tau}}3\varepsilon
(\gamma_m - <\gamma_\phi>) (y-\bar{y})\underbrace{(1-(y+\bar{y}))}_{\vert . \vert \leq 1} ds  \\
&\qquad   + \int_{\bar{\tau}_n}^{\bar{\tau}} {\Large(} \varepsilon^2\underbrace{h(y,w,\varepsilon, s)}_{\vert .
\vert \leq M}  + \mathcal{O}(\varepsilon^3) {\Large )}\, ds \big\vert\\
&\leq 3C \varepsilon_n  \int_{\bar{\tau}_n}^{\bar{\tau}} \vert \zeta(s) \vert ds  + \varepsilon_n^2
\int_{\bar{\tau}_n}^{\bar{\tau}} Mds+ \mathcal{O}(\varepsilon_n^3)) \\
&\leq 3C \varepsilon_n  \int_{\bar{\tau}_n}^{\bar{\tau}} \vert \zeta(s)
\vert ds + \varepsilon_n^2 M (\bar{\tau}- \bar{\tau}_n)+ \mathcal{O}(\varepsilon_n^3),
\end{split}
\end{equation}
for $\bar{\tau}\geq \bar{\tau}_{n+1}$, where we used the fact that 
  $|(\gamma_m-<\gamma_\phi>)|\leq C$ with $C>0$ constant. Applying Gronwall's Lemma (see Lemma
4.1.2 in~\cite{GuckHo}, or p. 37 in~\cite{CoLev}), results in
\begin{equation} \label{estimzeta}
\vert \zeta(\bar{\tau}) \vert \leq \frac{M}{3}\varepsilon_n (\exp(3C\varepsilon_n (\bar{\tau}- \bar{\tau}_n))-1)
+ \mathcal{O}(\varepsilon^2).
\end{equation}
Hence for $\bar{\tau}-\bar{\tau}_n\in [0,1/\varepsilon_n]$, i.e.
$\bar{\tau}\in [\bar{\tau}_n, \bar{\tau}_{n+1}]$, the inequality $\vert \zeta
(\bar{\tau})\vert \leq K \varepsilon_n $ holds for a positive constant $K$.
Letting $n$ go to infinity implies that  $\zeta$ tends to zero when
$\bar{\tau}$ goes to infinity. Therefore $y$ and $\bar{y}$  have the same
limit as $\bar{\tau}\to \infty$, i.e. 0 or 1 depending on the sign of the
quantity $\gamma_m - \la \gamma_\phi\ra $. Finally, recall that
$\Omega_{\phi}=y+\varepsilon w$; from the triangle inequality, and
$\varepsilon\rightarrow 0$ when $\bar{\tau}\rightarrow \infty$, it follows
that $\Omega_{\phi}$ also converges to 0 or 1 according to the sign of
$\gamma_m - \la \gamma_\phi\ra $. This completes the proof of the theorem for
the non-critical cases (i) and (ii) for which $\gamma_m - \la
\gamma_\phi\ra\neq 0$.

Let us now consider the critical case (iii), for which $\gamma_m = \la \gamma_\phi\ra
=2n/(n+1)$. In this case the right hand side of the averaged
equation~\eqref{averbary} vanishes. Furthermore, the evolution
equation~\eqref{eqfory} for $y$ becomes
\begin{equation} \label{eqforycrit}
\frac{dy}{d\bar{\tau}} =
\varepsilon^2  \Big\{ 3w (\gamma_m - \gamma_\phi )(1-2y)
+\frac1n (1-\varepsilon) (1+q)\Big\}
%&\quad - 3\frac{\partial w}{\partial y}   (\gamma_m - \la  \gamma_\phi \ra  ) y(1-y)
+ \mathcal{O}(\varepsilon^3)
%w (3(\gamma_m - \gamma_\phi (1-2y) ) +\frac1n (1-\varepsilon) (1+q)) - 3D_yw   (\gamma_m - < \gamma_\phi > ) y(1-y)
%\right)
\end{equation}
Let us first consider the average of the right hand side. We therefore define
\begin{equation}
g(y,w,\varepsilon, \bar{\tau})= 3w (\gamma_m - \gamma_\phi)(1-2y)
+\frac1n (1-\varepsilon) (1+q),
\end{equation}
and compute its average at $\varepsilon=0$, where $\Sigma_\dagger^2$ is a
periodic function and $\la 2 \Sigma_\dagger^2 \ra= \la \gamma_\phi \ra y$,
\begin{equation}
\begin{split}
\la g\ra (y,w)&= \frac1P \int_0^P g(y,w,0,\bar{\tau})d\bar{\tau}\\
&= 3w\underbrace{(\gamma_m - \la \gamma_\phi\ra)}_{=\, 0}(1-2y) +\frac1n \la 1+q \ra \\
%\frac1P \int_0^P(1+q)d\bar{\tau}\\
&=\frac{3}{2n} \frac1P \int_0^P(2\Sigma_\dagger^2+ \gamma_m (1-y))d\bar{\tau} \\
&= \frac{3}{2n} \left( \underbrace{(\gamma_m - \la \gamma_\phi\ra)}_{=\, 0}y + \gamma_m \right)\\
&=  \frac{3}{2n} \gamma_m
\end{split}
\end{equation}
Note that in the  critical case, $\frac{3}{2n}\gamma_m=\frac{3}{2n}\la
\gamma_\phi\ra =\frac{3}{2n}\frac{2n}{(n+1)}= \frac{3}{n+1}$. We therefore
obtain the following averaged system, expected to give the leading order
approximation:
\begin{subequations}\label{secondav}
\begin{align}
\frac{d\bar{z}}{d\bar{\tau}}&= \varepsilon^2 \frac{3}{n+1} \label{secondavzbar},\\
\frac{d \varepsilon}{d \bar{\tau}}&= -\frac{3}{n(n+1)} \varepsilon^2 (1- \varepsilon) .\label{secondsysavereps}
\end{align}
\end{subequations}
Again we study the dynamics on a Euler multiplied version of this system,
i.e. in the time variable ${\varepsilon} d/d\tau= d/d\bar{\tau}$:
\begin{subequations}\label{eulersecondav}
\begin{align}
\frac{d\bar{z}}{d\tau}&= \varepsilon \frac{3}{n+1} \label{eulersecondavzbar},\\
\frac{d \varepsilon}{d\tau}&= -\frac{3}{n(n+1)} \varepsilon (1- \varepsilon) .\label{eulersecondsysavereps}
\end{align}
\end{subequations}
The linearization at the line of fixed points $\{\varepsilon=0\}$
is
$$ \left(
\begin{array}{cc}
0 & \frac{3}{(n+1)} \\
0 &- \frac{3}{n(n+1)}
\end{array}
\right)
$$
This linearization admits one eigenvalue that is zero with a corresponding
eigenvector that is parallel to the line of fixed points, and a stable (i.e.
negative) eigenvalue $\lambda= -\frac{3}{n(n+1)}$, with a corresponding
eigenvector $(z=-n, \varepsilon=1 )$ pointing toward the inside of the
cylinder (i.e. the line of fixed points of the system~\eqref{eulersecondav}
is transversally hyperbolic).
%Hence there is
%for each fixed point on the line $\{\varepsilon=0\}$ a trajectory with
%positive initial $\varepsilon$ with a limit that is tangential to the above
%stable eigenvector. .
Hence each fixed point $(\bar{z}_0, 0)$ has a one-dimensional stable
manifold, i.e., there exists a trajectory $\bar{z}(\tau)$ that solves the
averaged system~\eqref{eulersecondav} with a positive initial $\varepsilon$
and converges to $(\bar{z}_0, 0)$ for each $\bar{z}_0$ as $\tau\rightarrow
\infty$. This asymptotic behavior is not affected by transforming the
equations back to the time variable $\bar{\tau}$.

A straightforward estimation of the term $\mathcal{O}(\varepsilon^3)$
provides bootstrapping sequences $\{\bar{\tau}_n\}_{n\in \mathbb{N}}$ and
$\{\varepsilon_n\}_{n\in \mathbb{N}}$, defined similarly as in the
non-critical case \eqref{bootstrap}. In other words, we obtain a
pseudo-trajectory $\{\Omega_\phi^n(\bar{\tau})\}$ of the original system
\eqref{Ophieps} with
\begin{equation}
\Omega_\phi^n(\bar{\tau}_n)=\bar{z}(\bar{\tau}_n),\qquad
\forall \bar{\tau} \in [\bar{\tau}_n, \bar{\tau}_{n+1}], \qquad  \vert \Omega_\phi^n(\bar{\tau}) -
\bar{z}(\bar{\tau})\vert < K \varepsilon_n,
\end{equation}
for a constant $K$. By regularity of the flow and compactness of the
cylinder, there is an initial data whose trajectory $\Omega_\phi(\bar{\tau})$
under the flow of the original equation \eqref{Ophieps} shadows the above
pseudo-trajectory in the following sense:
\begin{equation}
\forall n \in \mathbb{N}, \qquad \forall \bar{\tau} \in [\bar{\tau}_n,
\bar{\tau}_{n+1}], \qquad \vert \Omega_\phi^n(\bar{\tau}) -\Omega_\phi
(\bar{\tau})\vert < K \varepsilon_n.
\end{equation}
By the triangle inequality, $\vert \Omega_\phi^n(\bar{\tau})
-\bar{z}(\bar{\tau})\vert \to 0$ as $\bar{\tau}\to \infty$; therefore, for
each $\bar{z}_0\in [0,1]$, there exists a trajectory that is limiting to the
periodic trajectory at $\varepsilon=0$, characterized by $\Omega_\phi=
\bar{z}_0$.
%,
%combined with a bootstrapping argument that involves sequences
%$\{\bar{\tau}_n\}_{n\in \mathbb{N}}$ and $\{\varepsilon\}_{n\in \mathbb{N}}$,
%as in the previous cases, shows that $y-\bar{z}$ goes to zero as $\bar{\tau}$
%goes to infinity. Again by a triangular inequality, $\Omega_\phi- \bar{z}$
%also goes to zero.
Translating these results into the state space
of~\eqref{dynsys} concludes the proof of case (iii).
\end{proof}

Expressing these results for orbits with $0<\Omega_m<1$ and $0<T<1$ in terms
of formal global future attractors ${\cal A}_+$ of the global dynamical
system~\eqref{dynsys} leads to:\footnote{Loosely speaking, in dynamical
systems theory attractor behavior describes situations where a collection of
state space points evolve into a certain `attractor' region from which they
never leave. For a formal definition of a dynamical systems attractor, see
e.g.~\cite{waiell97}, and references therein.}
\begin{itemize}
\item[\rm{(i)}] ${\cal A}_+ = \bar{\bf S}|_{T=1,\Omega_m=0}$\!
    \qquad\qquad if \quad $\gamma_m > \frac{2n}{n+1}$,
\item[\rm{(ii)}] ${\cal A}_+ = \bar{\bf S}|_{T=1,\Omega_\phi=0} =
    \mathrm{FL}_1$ \quad if \quad $\gamma_m < \frac{2n}{n+1}$,
\item[\rm{(iii)}] ${\cal A}_+ = \bar{\bf S}|_{T=1}$ \qquad\qquad\qquad if
    \quad $\gamma_m = \frac{2n}{n+1}$.
\end{itemize}

\begin{figure}[ht!]
\begin{center}
\subfigure[The scalar field boundary]{\label{fig:Phi4Bound}
\includegraphics[width=0.45\textwidth]{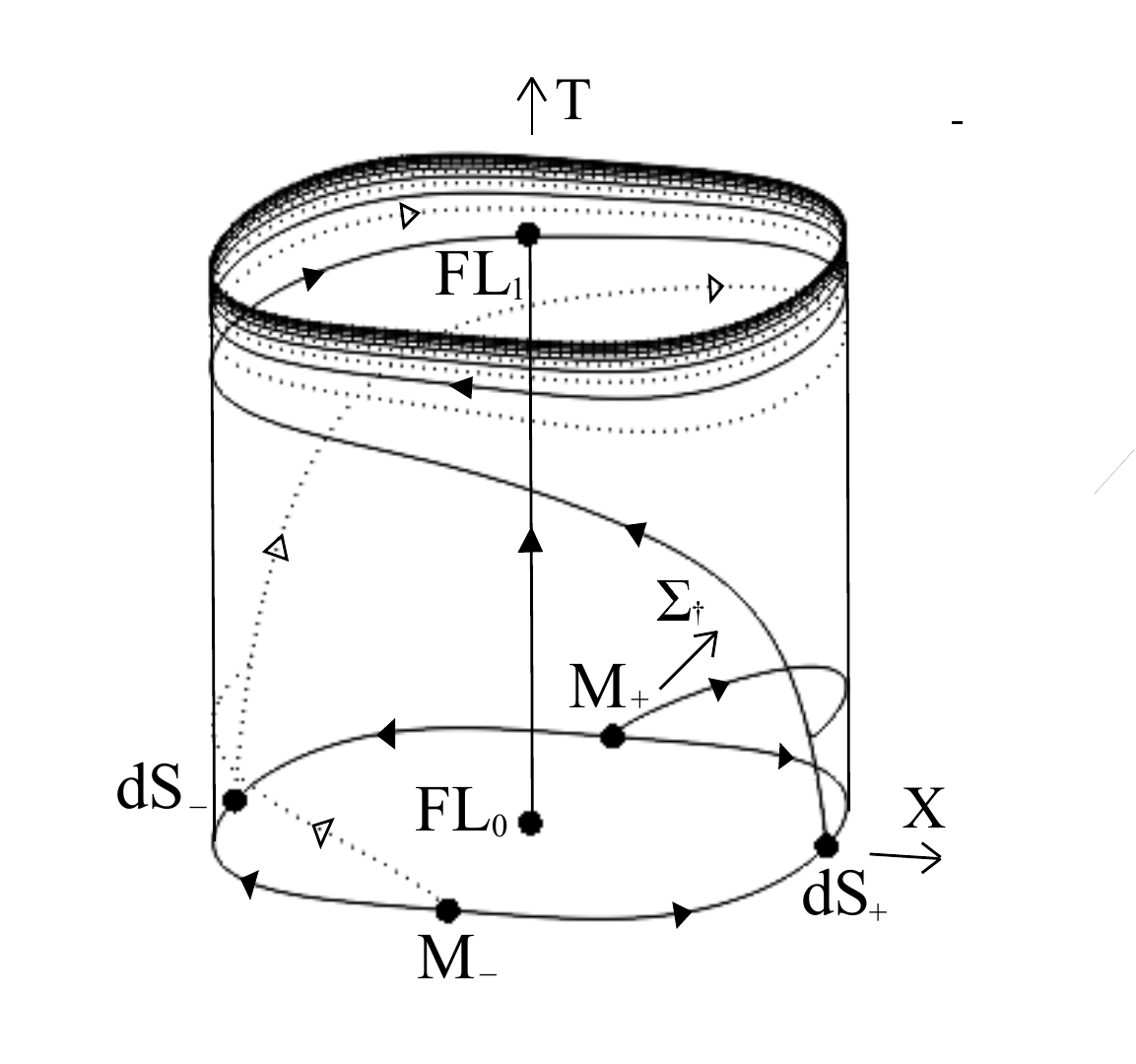}}
\subfigure[$\gamma_m=\frac{3}{2}$]{\label{fig:Phi4Radiation0}
\includegraphics[width=0.42\textwidth]{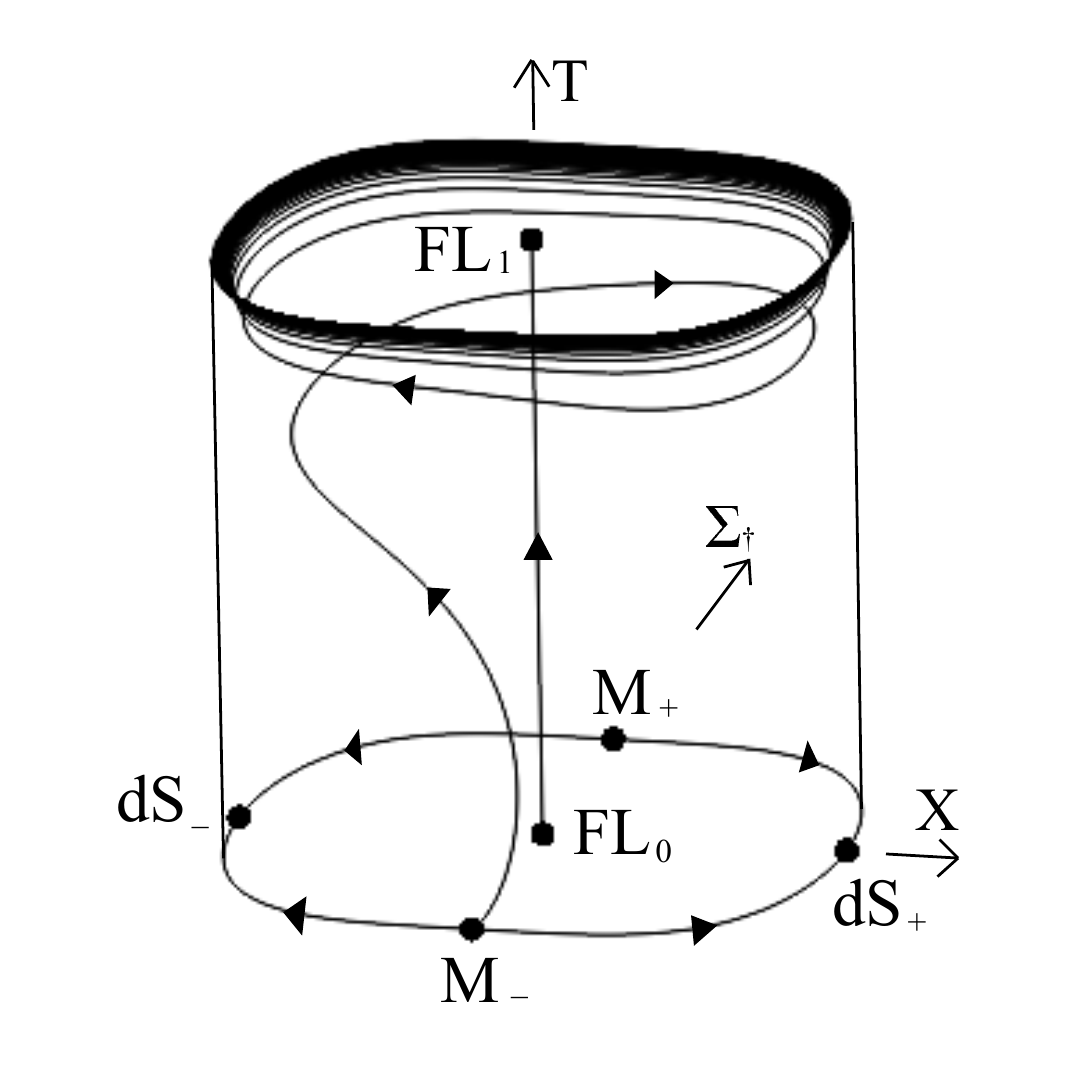}}
\subfigure[$\gamma_m=1$]{\label{fig:Phi4Dust}
\includegraphics[width=0.42\textwidth]{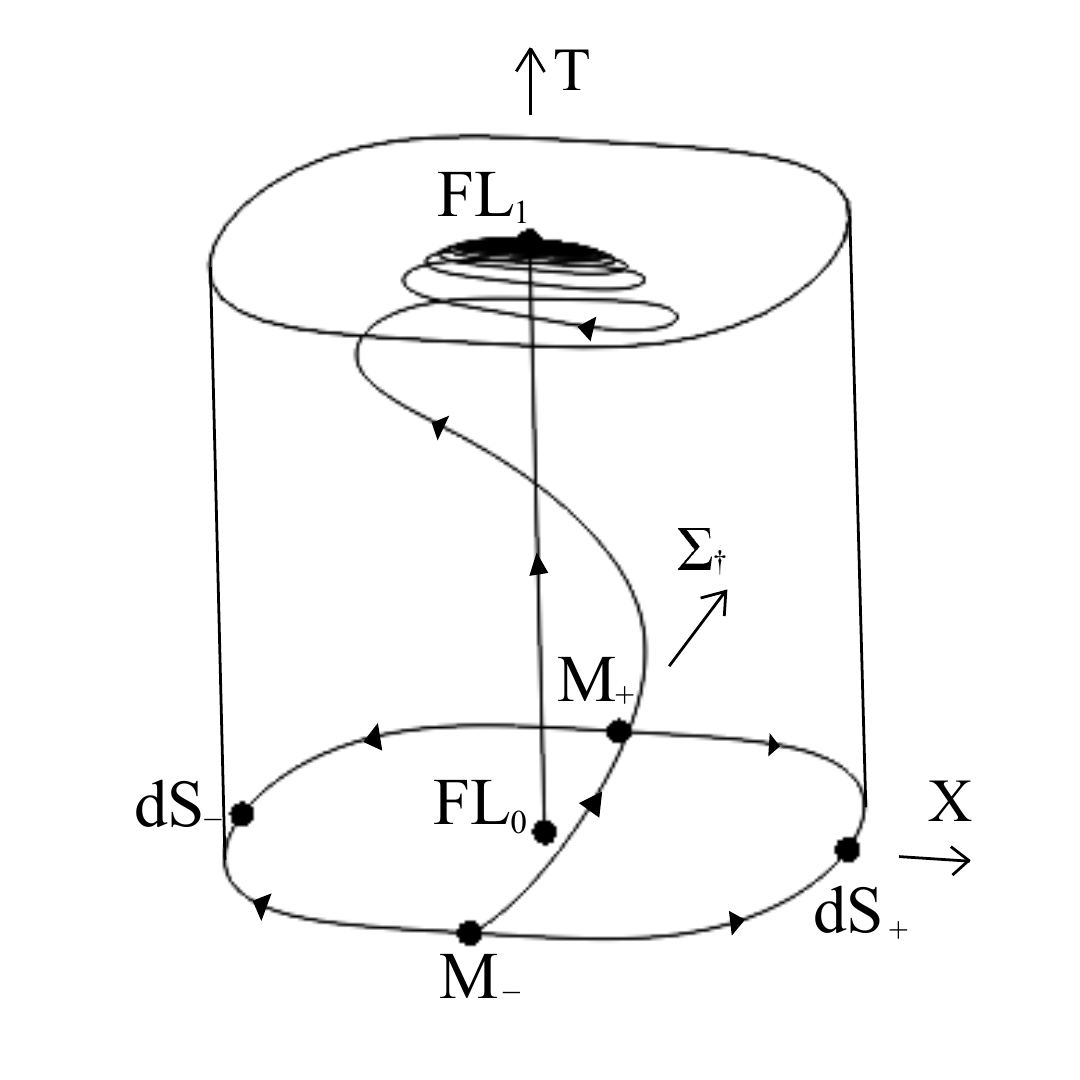}}
\subfigure[$\gamma_m=\frac{4}{3}$]{\label{fig:Phi4Radiation}
\includegraphics[width=0.42\textwidth]{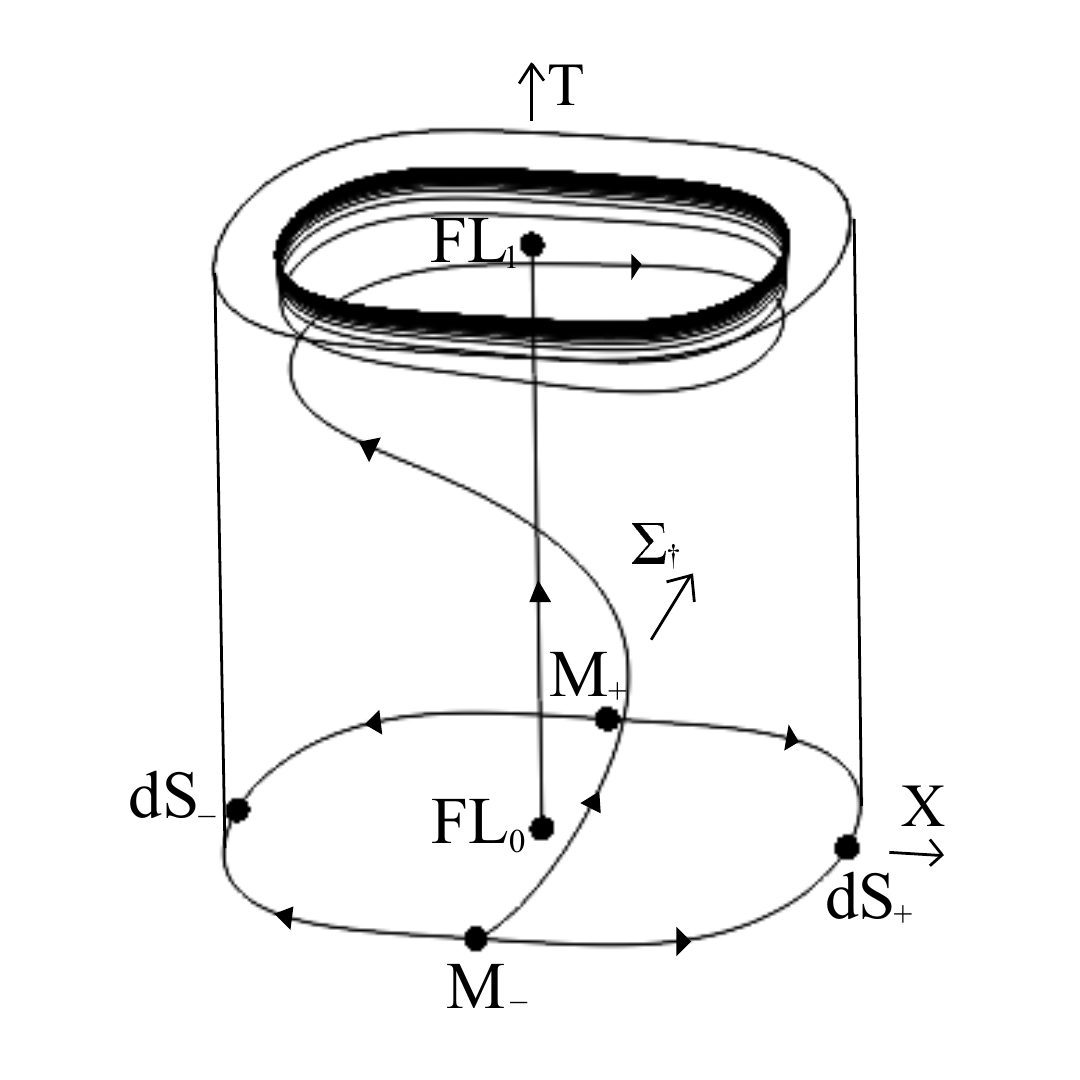}}
\end{center}
\caption{Solutions for the scalar field potential $V(\phi)=\frac{1}{4}(\lambda\phi)^4$
for various matter equation of states. The first picture describes a solution and
the attractor solutions on the scalar field boundary. The other figures depict
a typical solution that illustrates the behavior at late times for cases
(i) (i.e., $\gamma_m > \frac{2n}{n+1}$), (ii) (i.e., $\gamma_m < \frac{2n}{n+1}$),
and (iii) (i.e., $\gamma_m = \frac{2n}{n+1}$).}\label{fig:LateTime}
\end{figure}

The above results are illustrated by the numerical examples that are depicted
in Figure~\ref{fig:LateTime}.
\subsubsection*{Physical Interpretation}

The present results have physical consequences. The solutions approach
$(H,\phi,\dot{\phi},\rho_m) =(0,0,0,0)$ toward the future, which is the
Minkowski solution of the system~\eqref{Hphieq}. The Minkowski solution
admits an 11-dimensional homothety group and is therefore an example of a
self-similar spacetime. However, the asymptotic future state for solutions in
case (i), with $\gamma_m > \frac{2n}{n+1}$ and $\Omega_\phi>0$, is the
periodic orbit at $\Omega_\phi = 1 = T$, and hence, since the future
asymptotic behavior is described by a limit cycle, it follows that the
deceleration parameter $q$ oscillates toward the future (this is also
obviously true for the orbits on the subset $\Omega_\phi=1$). The solutions
thus approach the Minkowski spacetime in a foliation that is induced by the
present models in a manner where the self-similar nature of the Minkowski
spacetime is not manifest and thus we say that these models exhibit
\emph{future asymptotic manifest self-similarity
breaking}.\footnote{Asymptotic (continuous) manifest self-similarity here
implies that physical geometrical scale-invariant observables such as the
deceleration parameter $q$ take asymptotic constant values, but since the
future attractor is a limit cycle it follows that $q$ is asymptotically
oscillating, i.e., asymptotic manifest self-similarity is broken. For another
example of future asymptotic self-similarity breaking in cosmology,
see~\cite{waietal99}. Note that asymptotic manifest self-similarity is a
complicated issue in the present asymptotic Minkowski case due that the only
metrics with no conformal scalars (and hence the only admitting a local
conformal group not conformally isometric) are either conformal to the plane
wave metric with parallel rays or conformally Minkowski,
see~\cite{def75,ear74}.}

Physically, case (ii), for which $\gamma_m < \frac{2n}{n+1}$, has the
simplest asymptotic regime. The future state for this case when $\Omega_m>0$
is the fixed point $\mathrm{FL}_1$ for which $q = \frac12(3\gamma_m-2)$. This
state describes the flat perfect fluid model with equation of state parameter
$\gamma_m$. Since this model is self-similar, the solutions asymptotically
approach the Minkowski spacetime in a manifest self-similar manner and we
therefore say that the present class of models are \emph{future
asymptotically} (manifestly) \emph{self-similar}.

Case (iii), where $\gamma_m = \frac{2n}{n+1}$, implies that
(with the exception of the perfect fluid solution with $\Omega_\phi=0$) all
solutions oscillate asymptotically toward the future and thus this case is
also characterized by future asymptotic manifest self-similarity breaking.
Since asymptotic manifest self-similarity, and breaking thereof, plays an
important role in physics as a whole, it is of considerable interest to note
the above features.

Furthermore, recall that the case $\gamma_m = \frac23$ is equivalent to a
pure scalar field case in an open FLRW model. As a consequence \emph{all}
monomial potentials (with $n\geq 1$) for these models lead to that the future
end state is given by the fixed point $\mathrm{FL}_1$, which in this context
is associated with a Milne state.

We are now finally in a position to describe the global solution space and
its features. There is a 1-parameter set of solutions that originate from the
fixed point $\mathrm{FL}_0$, corresponding to an initial perfect fluid
dominated state, into the state space ${\bf S}$; two equivalent 2-parameter
sets of solutions that originate from each of the equivalent fixed points
$\mathrm{M}_\pm$, corresponding to a self-similar initial massless scalar
field dominated state, where two equivalent 1-parameter subsets, belonging to
each of these sets of solution, reside on the scalar field boundary
$\Omega_\phi = 1$; finally, there are two equivalent attractor solutions
residing on $\Omega_\phi = 1$ that originate from each of the fixed points
$\mathrm{dS}_\pm$, thus exhibiting an initial de Sitter state in the sense
previously discussed. The future state of all orbits in ${\bf S}$ resides on
the subset $T=1$, where the precise asymptotic location on $T=1$ depends on
the relation between the monomial potential exponent $2n$ and the equation of
state parameter $\gamma_m$ according to Theorem~\ref{theorem}.

In all three cases (i), (ii), and (iii) there exists an open set of solutions
that is close to the attractor solutions at some intermediate stage of their
evolution, but there also exists an open set of solutions that is not close
to the attractor solutions. Furthermore, although there exists an open set of
solutions that is close to the attractor solutions at late times in case (i)
this is not true for case (ii) where each (equivalent) attractor solution
only acts as a kind of `saddle.' In addition, note that the variables $T,
\Sigma_\dagger, X$ can be locally expressed in terms of the physical
quantities $H, q, \Omega_m$. By imposing a Euclidian measure on the space
described by $H, q, \Omega_m$ one finds that `most' of the evolution of
`most' solutions is not governed by the `attractor' solution; the addition of
a degree of freedom, in this case associated with a perfect fluid, has
aggravated the situation for arguments that attempt to establish that
attractor solutions in some sense are attractors.\footnote{For further
discussions about the meaning of `attractor solutions' and measures,
see~\cite{alhugg15} and the recent papers by Remmen and
Carroll~\cite{remcar13,remcar14} and by Corichi and Sloan~\cite{corslo14},
and references therein.} Nevertheless, attractor solutions are likely to
continue to generate considerable attention, and thus there is a need to
describe them accurately, which is the topic of the next section.

%%%%%%%%%%%%%%%%%%%%%%%%%%%%%%%%%%%%%%%%%%%%%%%%%%%%%%%%%%%%%%%%%%
\section{Attractor solution approximants}\label{sec:attrappr}
%%%%%%%%%%%%%%%%%%%%%%%%%%%%%%%%%%%%%%%%%%%%%%%%%%%%%%%%%%%%%%%%%%

In this section we will introduce and compare several analytical
approximation schemes that describe the attractor solution. Since the two
equivalent attractor solutions are just the center manifolds of
$\mathrm{dS}_\pm$, we begin with a direct approach of approximately obtaining
these center manifolds by means of center manifold theory. For presentations
of center manifold analysis, see e.g.,~\cite{car81,cra91}, and for
applications in cosmology, e.g.,~\cite{ren02,bohetal12,alhugg15}.

%, thus exploring
%how the transformation between these variables affect approximants.
%investigate and compare different approximants to the inflationary attractor
%solution on the scalar field boundary ${\bf S}|_{\Omega_m=0}$. We begin by
%considering slow-roll approximants which are the basis for much work
%concerning predictions for inflationary quantities, such as the tensor to
%scalar ratio or the spectral index of primordial perturbations~\cite{Muk05}.
%We then construct center manifold approximants, as introduced
%in~\cite{AU2014}, as well as other approximants.

%---------------------------------------------------------------
\subsection{Center manifold approximants}
%---------------------------------------------------------------

The equivalent center manifolds of $\mathrm{dS}_\pm$ reside on the scalar
field subset $\Omega_m=0$ and therefore it suffices to study the equations on
this subset. In~\cite{alhugg15} it was shown that for quadratic potentials
the center manifold expansion using the bounded system~\eqref{dynsysB} for
$n=1$ resulted in a larger range than the expansion based on the unbounded
system~\eqref{dynsysBearly}. However, it was also shown that by using these
expansions to produce so-called Pad{\'e} approximants resulted in much better
approximations, both as regards accuracy and range. Moreover, the Pad{\'e}
approximants for the system~\eqref{dynsysBearly} gave the same expressions as
the Pad{\'e} approximants obtained from~\eqref{dynsysB} (or, more accurately,
they produced the converging subset of Pad{\'e} approximants associated
with~\eqref{dynsysB}), but in a simpler form. Thus the most useful results
from center manifold analysis are obtained by performing such an analysis for
the unbounded system~\eqref{dynsysBearly}, followed by an introduction of
Pad{\'e} approximants, instead of using~\eqref{dynsysB}.
%This system also has the advantage that the
%independent variable can be interpreted as the number of $e$-folds, something
%that will be further used below.
%These aspects provide motivation for
%carrying out the center manifold analysis using the
%system~\eqref{dynsysBearly} instead of~\eqref{dynsysB}.

Due to the discrete symmetry, we can, without loss of generality, restrict
the center manifold analysis to the fixed point $\mathrm{dS}_+$ at
$\theta=0=\tilde{T}$. A linearization of the system~\eqref{dynsysBearly} at
this fixed point yields
\begin{subequations}\label{linearanalysis}
\begin{align}
E^s &= \{(\tilde{T},\theta)|\tilde{T}=0\},\\
E^c &= \{(\tilde{T},\theta)|\sqrt{n}\tilde{T} + 3\theta = 0\},
\end{align}
\end{subequations}
where $E^s$ and $E^c$ denote the tangential stable and center subspaces,
respectively. Since the tangential center subspace is given by
$\sqrt{n}\tilde{T} + 3\theta =0$, we introduce
\begin{equation}
v=\tilde{T} + \frac{3}{\sqrt{n}}\theta
\end{equation}
as a new variable that replaces $\theta$ in order to study the center
manifold $W^c$, with the tangent space $E^c$ at $(\tilde{T},v) = (0,0)$. As
follows from~\eqref{dynsysBearly}, this leads to the transformed system
\begin{subequations}\label{dynsysv}
\begin{align}
\frac{d\tilde{T}}{d\tau} &= \frac{3}{n}\tilde{T}\left(1-\cos^{2n}\theta\right), \label{Tveq}\\
\frac{dv}{d\tau} &= \frac{3}{\sqrt{n}}\left(\frac{1}{\sqrt{n}}\tilde{T}\left(1-\cos^{2n}\theta\right)
- \tilde{T}F(\theta) - \frac{3}{2n}F^2(\theta)\sin 2\theta\right) ,\label{vveq}
\end{align}
\end{subequations}
where $\theta = \frac{\sqrt{n}}{3}(v-\tilde{T})$. The linearization of
eq.~\eqref{vveq} yields $\frac{dv}{d\bar{\tau}} = - 3v$, while
eq.~\eqref{Tveq} only has higher order terms. The center manifold $W^c$ can
be obtained as the graph $v = h(\tilde{T})$ near $(\tilde{T},v) = (0,0)$
(i.e., use $\tilde{T}$ as an independent variable), where $h(0) =0$ (fixed
point condition) and $\frac{dh}{d\tilde{T}}(0) =0$ (tangency condition).
Inserting this relationship into eq.~\eqref{dynsysv} and using $\tilde{T}$ as
the independent variable leads to
\begin{equation}\label{htildeT}
\frac{1}{\sqrt{n}}\tilde{T}\left(1-\cos^{2n}\theta\right)\left(\frac{dh}{d\tilde{T}}-1\right) +
\tilde{T}F(\theta) + \frac{3}{2n}F^2(\theta)\sin2\theta = 0,
\end{equation}
where $\theta = \frac{\sqrt{n}}{3}(h(\tilde{T})-\tilde{T})$.

Solving the above nonlinear ordinary differential equation amounts to finding
the attractor solution, which for the present class of problems does not seem
likely to be possible. Instead the equation can be solved approximately by
representing $h(\tilde{T})$ as a formal truncated power series and by Taylor
expanding the expressions involving $\theta$ (and subsequently replace
$\theta$ with its power series expression in $\tilde{T}$), which makes it
possible to algebraically solve for the coefficients in the formal series.
Before doing this, it is useful to note that all coefficients in the above
equation are odd in terms of $\tilde{T}$ and $\theta$, and as a consequence
the power series for $h$ consists of odd powers of $\tilde{T}$, since it is
only then $\frac{dh}{d\tilde{T}}$ results in even powers (odd powers for
$\frac{dh}{d\tilde{T}}$, and hence even powers for $h$, are clearly zero due
to the above properties). Furthermore, it follows from the tangency condition
that the series for $h$ in odd powers of $\tilde{T}$ begins with a cubic
term. Hence
\begin{equation}
h(\tilde{T}) = \sum_{i=1}^n a_i\tilde{T}^{2i + 1} + \mathcal{O}(\tilde{T}^{2n+3}) =
\tilde{T}\sum_{i=1}^n a_ix^{i} + \mathcal{O}(\tilde{T}^{2n+3}) \qquad \text{as}\qquad \tilde{T}\rightarrow 0,
\end{equation}
where the series for $h(T)$ is truncated at some chosen order and where we
have introduced $x=\tilde{T}^2$. Inserting this into a Taylor expansion of
eq.~\eqref{htildeT} and algebraically solving for the coefficients leads to
\begin{equation}\label{CM_Expansion}
\begin{split}
\theta &\approx - \frac{\sqrt{n}}{3}\tilde{T}f(x),\\
f(x) &= 1 + \frac{n}{108}\left[(3n-7)x + \frac{n}{360}(65n^2 - 570n + 1081)x^2\right],
\end{split}
\end{equation}
where we have chosen to truncate the series for $\theta$ at 5th order in
$\tilde{T}$.

To improve the range and accuracy of the above approximation for the
attractor solution we construct the $[1/1]_{f}(x)$ Pad{\'e} approximant,
which leads to the following approximate expression for $\theta$:\footnote{A
Pad{\'e} approximant of order $(m,n)$ of a function $f(x)$, denoted by
$[m/n]_f(x)$, is associated with a truncated Taylor series $f \approx c_0 +
c_1 x + c_2 x^2 + \cdots + c_{m+n}x^{m+n}$ and given by the polynomials
$P_m(x) = p_0 + p_1 x + p_2 x^2 + \cdots+p_m x^m$ and $Q_n(x) = q_0 + q_1 x +
q_2 x^2+\cdots+q_n x^n$ according to $[m/n]_f(x) = \frac{P_m(x)}{Q_n(x)}$,
such that $Q_n(x)(c_0 + c_1 x + c_2 x^2 + \cdots + c_{m+n}x^{m+n}) = P_m(x)$,
where coefficients with the same powers of $x$ are equated up through $m+n$.
For details and examples, see~\cite{bak75,kal02,sinetal03,alhugg15} and
references therein.}
\begin{equation}\label{CM_Pade}
\theta \approx - \frac{\sqrt{n}}{3}\tilde{T}[1/1]_{f}(x), \qquad
[1/1]_{f}(x) = \frac{1-\frac{n}{36(3n-7)}
\left(\frac{7}{2}n^2-43n+\frac{2753}{30}\right)x}{1-\frac{n}{36(3n-7)}
\left(\frac{13}{2}n^2-57n+\frac{1081}{10}\right)x} .
\end{equation}
When this is subsequently expressed in $T$, i.e., $\tilde{T}= T/(1-T)$ and
$x=T^2/(1-T)^2$, this yields a curve $\theta(T)$ which approximates the
attractor solution in the state space ${\bf S}$ on the boundary $\Omega_m=0$.
The case $n=1$ was dealt with in  Figures 5 and 14 in~\cite{alhugg15}, where
these approximations curves for the attractor solution, as well as those
obtained by considering higher order expansions, were compared with the
numerically computed attractor solution. To obtain explicit curves in our
state space picture we need to specify $n$. To avoid details, we in this
paper only compare the approximate solution curves with the numerically
computed attractor solution for some representative values of $n$ for the
above $[1/1]_{f}(x)$ Pad{\'e} approximant in subsection~\ref{sec:comp} below;
although it should be pointed out that higher order Pad{\'e} approximants
give better, although more complicated, results.

Ref.~\cite{alhugg15} also illustrates that nonlinear transformations might
have two important features. First, they affect analytical or numerical
computations and therefore a suitable choice of variables can make a problem
more tractable. Second, they affect approximations and a suitable choice of
variables can lead to better approximations for solutions like the attractor
solution. We will therefore next consider an approximation scheme based on
the variables $\Sigma_\dagger$ and $X$ instead of $\theta$.
%(which is the one natural leading to the center manifold analysis).

%---------------------------------------------------------------
\subsection{Series expansion approximants}\label{sec:series}
%---------------------------------------------------------------

As in the previous case we use a system adapted to the dynamics at early
times and we therefore use $\tilde{T}$, but $\theta$ is replaced with
$\Sigma_\dagger$ and $X$, i.e., we consider the system~\eqref{dynsysearly}.
We are interested in finding new approximations for the attractor solution,
and inspired by the result in eq.~\eqref{CM_Expansion} we assume that
$\Sigma_\dagger$ and $X$ can be written as formal truncated series in
$\tilde{T}$. These series are subsequently inserted in~\eqref{dynsysearly},
but since we are interested in the attractor solution, which resides on the
$\Omega_m=0$ boundary, we also require that the constraint $\Sigma_\dagger^2
+ X^{2n}=1$ is satisfied. The analysis is simplified by noting that it
follows from the system~\eqref{dynsysearly} that $\Sigma_\dagger$ must have
only odd terms while $X$ have only even terms in their series expansions in
$\tilde{T}$. Moreover, since we choose, without loss of generality, to
consider the solution that originates from the $\mathrm{dS}_+$ fixed point it
follows that to lowest order $\Sigma_\dagger =0, X=1$. Algebraically solving
for the coefficients, and writing $\tilde{T}^2=x$ as before, leads to
\begin{subequations}\label{SigmaXTildeT}
\begin{align}
\Sigma_\dagger &\approx -\frac{n}{3}\tilde{T}f(x),\qquad
f(x) = 1-\frac{n}{18}x + \frac{n^2}{648}(17-6n)x^2,\\
X &\approx 1 -\frac{n}{18}x + \frac{n^2}{648}(5-2n)x^2,\label{XT}
\end{align}
\end{subequations}
where we have chosen to truncate the series for $f(x)$ and $X(x)$ at 2nd
order in $x$.

To improve the range and accuracy we calculate the $[1/1]_{f}(x)$ and
$[1/1]_{X}(x)$ Pad{\'e} approximants, which yield the following approximate
expressions
\begin{subequations}\label{SigmaXPade}
\begin{align}
\Sigma_\dagger &\approx  -\frac{n}{3}\tilde{T}\left[\frac{1 + \frac{n}{12}(5-2n)x}{1 + \frac{n}{36}(17-6n)x}\right],\label{seriesSigma}\\
X &\approx \frac{1+\frac{n}{36}\left(3-2n\right)x}{1+\frac{n}{36}\left(5-2n\right)x}.\label{XPade}
\end{align}
\end{subequations}
Note that for small values of $\tilde{T}$ in the neighborhood of
$\mathrm{dS}_+$ we obtain $X= 1 - \frac{n}{18}\tilde{T}^2 = \cos\theta
\approx 1 - \frac12\theta^2$, which gives $\theta =
-\frac{\sqrt{n}}{3}\tilde{T}$, which is just the tangency condition for the
center manifold of $\mathrm{dS}_+$. Replacing $\tilde{T}$ and $x$ with $T$
in~\eqref{SigmaXPade} yield two curves in the state space ${\bf S}$ on the
scalar field subset $\Omega_m=0$, one for the approximation of $X$ and one
for the approximation of $\Sigma_\dagger$, both approximating the attractor
solution.

To obtain accurate numerical results for the attractor solution it is
preferable to use the unconstrained system~\eqref{dynsysB} rather
than~\eqref{dynsys} subjected to the constraint $\Sigma_\dagger^2 +
X^{2n}=1$, especially when $\gamma_m < \frac{2n}{n+1}$ since the constraint
surface then becomes unstable for sufficiently large $T$. The above results
are translated into the corresponding $\theta(T)$ curves via $\theta =
-\arccos{X}$ (the minus sign is due to that $\theta<0$ and taking the default
branch) and implicitly via
$F(\theta)\sin{\theta}=-\frac{n}{3}\left(\frac{T}{1-T}\right)f$, where
$X(x(T))$ and $f(x(T))$ are obtained by replacing $x$ with $T$
in~\eqref{SigmaXTildeT} and~\eqref{SigmaXPade}, which lead to approximations
that subsequently can be compared with the numerically computed attractor
solution. Below, in subsection~\ref{sec:comp} we for brevity only discuss the
Pad{\'e} approximant given in eq.~\eqref{seriesSigma} for some representative
values of $n$.

As a final remark, note that the lowest order expansion just gives $X=1$,
which is equivalent to $\Sigma_\dagger=0$, since $X^{2n} = 1 -
\Sigma_\dagger^2$ on the scalar field boundary. The condition
$\Sigma_\dagger=0$ and $X=1$ describes a straight vertical line in $\bar{\bf
S}$, corresponding to $q=-1$, originating from $\mathrm{dS}_+$, which apart
from $\mathrm{dS}_+$ is not a good approximation for the attractor solution.
This is in contrast with the previous center manifold expansion, which is
given by a unique curve $\theta(T)$, which to all orders in the limit of
small $T$ is tangent to the center manifold.

Next we turn to approximations based on the slow-roll approximation and its
extensions.

%---------------------------------------------------------------
\subsection{Slow-roll based approximants}

We here extend the work in~\cite{lidetal94} and~\cite{alhugg15} to general
$n$ and to higher order by approximating $H^2(\phi)$ with so-called slow-roll
Hubble expansions, and illustrate this class of approximations in our global
state space picture. To facilitate comparison with the inflationary
literature, we initially keep the coupling constant $\kappa= 8\pi G=8\pi
m^{-2}_\mathrm{Pl}$ in the slow-roll Hubble expansion formulas below. The
approach developed in~\cite{lidetal94} is based upon a hierarchy of slow-roll
parameters, the first being
\begin{equation}
\epsilon_{H}= 3\left(\frac{\sfrac12\dot{\phi}^2}{\sfrac12\dot{\phi}^2 + V(\phi)}\right) = 1 + q, \qquad
\eta_{H} = -3\left(\frac{\ddot{\phi}}{3H\dot{\phi}}\right),
\end{equation}
which are assumed to be small. These parameters allow one to produce a
truncated expansion for the Hubble variable in terms of $\phi$, which for the
present monomial scalar field potential is given by (following the
prescription given in~\cite{lidetal94}, which we refer to for details):
\begin{equation}\label{Hslowroll}
\begin{split}
3H^2 &\approx \frac{\kappa}{2n}(\lambda\phi)^{2n}f(y) = \kappa V(\phi)f(y),\\
f(y) &= 1 + y + \left(1-\frac{2}{n}\right)y^2 + \left(1-\frac{3}{n}\right)^2y^3 +
\left(\left(1-\frac{4}{n}\right)^3+\frac{2}{n^3}\right)y^4 + \mathcal{O}\left(y^5\right)
\end{split}
 %\left[1+\frac{2n^2}{3\kappa\phi^{2}}+\frac{4n^3 (n-2)}{9\kappa^{2}\phi^4}
 %     +\frac{8n^4 (n-3)^2}{27\kappa^{3}\phi^6}+\frac{16n^5 ((n-4)^3+2)}{81\kappa^{4}\phi^8}
 %                        1+x+\left(1-\frac{2}{n}\right)x^2+\left(1-\frac{3}{n}\right)^2x^3
 %                        +\left(\left(1-\frac{4}{n}\right)^3+\frac{2}{n^3}\right)x^4+\mathcal{O}\left(x^5\right)\right],
\end{equation}
where we have found it convenient to define a quantity $y$ according to
\begin{equation}
y = \frac{2n^2}{3\kappa\phi^2}.
\end{equation}
Note that the 2nd (3rd) order term is zero in~\eqref{Hslowroll} when $n=2$
($n=3$).
% We have chosen to go to higher order for this expansion than for the
% previous approximants since it is of some interest to note that there then
% are differences between the Canterbury approximants used by Liddle {\it et
% al.}~\cite{lidetal94} and the Pad{\'e} approximants associated with the above
% series for $f(y)$, as will be seen below.
The above series expansion contains no parameters and thus it can only
describe a single solution, moreover, for the series expansion to make sense
requires that $y$ is small and hence that $\phi$ is large, which leads to
that $H$ is large, i.e., for the present models the Hubble slow-roll
expansion attempts to approximately describe the dynamics at early times for
a certain solution, but which one?

In terms of $T$ and $X$, and therefore in $T$ and $\theta$, the definitions
in eq.~\eqref{vardef1} lead to that the Hubble expansion~\eqref{Hslowroll}
can be written on the form (from now on we set $\kappa = 1$)
\begin{equation}\label{yxt}
1 \approx X^{2n}f(y) = \cos^{2n}\theta f(y) \quad \text{where} \quad
y = \left(\frac{n\tilde{T}}{3X}\right)^2 = \left(\frac{n\tilde{T}}{3\cos\theta}\right)^2,
\end{equation}
(as usual, $\tilde{T} = T/(1-T)$) where each order yields an implicit
relation for a curve on the scalar field boundary, although it can be used to
provide a series expansion for $X$ in terms of $\tilde{T}$ for small
$\tilde{T}$ and hence for small $T$.

Note that just as in the previous approximation scheme, the zeroth order
Hubble expansion just gives $X^{2n}=1$, which, as stated earlier,
% is equivalent to $\Sigma_\dagger=0$,
% since $X^{2n} = 1 - \Sigma_\dagger^2$ on the scalar field boundary. The
% condition $\Sigma_\dagger=0$ describes straight vertical lines in $\bar{\bf
% S}$, corresponding to $q=-1$, originating from $\mathrm{dS}_\pm$, which apart
% from $\mathrm{dS}_\pm$
is not a good approximation for the attractor solution. Nevertheless, this
approximation is what is sometimes referred to as the slow-roll Hubble
approximation, see e.g.~\cite{salbon90}. At first order, when $f(y) = 1+y$,
eq.~\eqref{yxt} results in
\begin{equation}\label{firstHy}
\frac{1-X^{2n}}{X^{2(n-1)}} = \frac{1-\cos^{2n}\theta}{\cos^{2(n-1)}\theta} = \left(\frac{n}{3}\,\tilde{T}\right)^2.
\end{equation}

Without loss of generality, consider the neighborhood of $\mathrm{dS}_+$;
then the above expression yields the limit (beyond the $\mathrm{dS}_+$ point
itself)
\begin{equation}
\theta \approx - \frac{\sqrt{n}}{3}\,\tilde{T} \approx - \frac{\sqrt{n}}{3}\,T,
\end{equation}
which is the tangency condition for the center submanifold of
$\mathrm{dS}_+$. Thus all curves associated with eq.~\eqref{Hslowroll}
originate from $\mathrm{dS}_\pm$; furthermore, all curves associated with
orders larger than zero are tangential to the attractor solutions in the
small $T$-limit toward $\mathrm{dS}_\pm$.

% Comparing with the numerical solution for the attractor solution we find that
% the Hubble slow-roll expansions improve with increasing order when $q$ is
% close to $-1$ and the slow-roll conditions are well satisfied, but this is no
% longer the case when $q$ becomes larger. In this case the results depend on
% $n$ in a rather unpredictable pattern, e.g., the 4th order expansion gives
% the worst result at $q=0$.

To improve the range and rate of convergence one can compute Pad\'e
approximants for the Hubble expansion for $f(y)$, i.e.,
\begin{equation}
3H^2 \approx \frac{1}{2n}(\lambda\phi)^{2n}[L/M]_f(y) = V(\phi)[L/M]_f(y),
\end{equation}
where
\begin{subequations}\label{HSRPADEs}
\begin{align}
[1/1]_f(y)&=\frac{1+\frac{2}{n}y}{1-\left(1-\frac{2}{n}\right)y}, \\
[2/2]_f(y)&=\frac{1-\frac{5}{2n}-
\left(1-\frac{11}{n}+\frac{22}{n^2}\right)y+\frac{8}{n^2}
\left(1-\frac{43}{16n}\right)y^2}{1-\frac{5}{2n}-2\left(1-\frac{27}{4n}+
\frac{11}{n^2}\right)y+\left(1-\frac{9}{n}+\frac{25}{n^2}-\frac{43}{2n^3}\right)y^2}.
\end{align}
\end{subequations}
In Ref.~\cite{alhugg15}, these approximations have been compared with the
numerically computed attractor solutions in detail for $n=1$, see Fig. 7 and
Table 3 of~\cite{alhugg15}.\footnote{In~\cite{alhugg15} the authors used the
slow-roll parameters to introduce so-called Canterbury approximants in order
to improve the range and accuracy of the approximations. We find that
Canterbury approximants give more cumbersome and poorer approximations than
the corresponding Pad{\'e} approximants when they differ (to lowest order
they agree with each other).}

Let us now turn to what is referred to as \emph{the slow-roll approximation}
in e.g.~\cite{salbon90}. This approximation is obtained by inserting the
lowest order Hubble expansion approximant $H=\sqrt{V(\phi)/3}$ into
\begin{equation}\label{dotphiHphi}
\dot{\phi} = - 2\frac{\partial H}{\partial \phi},  %=-\sqrt{\frac{2n}{3}}\lambda^{n}\phi^{n-1}\frac{f(x)-\frac{x}{n}\frac{df}{dx}}{\sqrt{f(x)}}.
\end{equation}
which yields
\begin{equation}
\dot{\phi} \approx -\sqrt{\frac{2n}{3}}\lambda^n\phi^{n-1}
\end{equation}
in the present case with a monomial potential. Expressed in terms of the
variables $\Sigma_\dagger$ and $X$, this gives
\begin{equation}\label{slow-roll}
\Sigma_\dagger \approx -\frac{n}{3}X^{n-1}\tilde{T},
\end{equation}
which when squared leads to
\begin{equation}
\frac{\Sigma_\dagger^2}{X^{2(n-1)}} = \frac{1-X^{2n}}{X^{2(n-1)}}
= \frac{1-\cos^{2n}\theta}{\cos^{2(n-1)}\theta} \approx \left(\frac{n}{3}\tilde{T}\right)^2.
\end{equation}
Note that this expression coincides with the first order Hubble expansion
approximation given in eq.~\eqref{firstHy}, and thus the slow-roll
approximation is an approximation to the attractor solution, since it yields
a curve that is tangential to the center manifold in the limit of small $T$.
Thus, \emph{apart from the zeroth order approximation, Hubble expansion
slow-roll approximations, as well as all slow-roll Hubble expansion based
approximants, are approximations for the attractor solution}.

The slow-roll approximation can be generalized by inserting higher order
Hubble approximants into $f(y)$. A complication arises since different
approximations can be obtained by multiplying expressions by various
combinations of $1 \approx X^{2n}f(y)$ (they are all tangential to the center
manifolds in the limit of small $T$); we here choose a combination that
yields the above slow-roll approximation (which was chosen to coincide with
the first order expression in~\eqref{firstHy}) when setting $f(y)=1=g(y)$
below:
\begin{equation}\label{SR_SIGMA_X}
\begin{split}
\Sigma_{\dagger} &\approx -\frac{n}{3}X^{n-1}\tilde{T}\frac{g(y)}{\sqrt{f(y)}},\\
g(y) &= f(y) - \frac{y}{n}\frac{df}{dy}\\
&= 1 + \left(1-\frac{1}{n}\right)y+\left(1-\frac{2}{n}\right)^2y^2 + \left(1-\frac{3}{n}\right)^3y^3\\
& \quad +\left(\left(1-\frac{4}{n}\right)^3 + \frac{2}{n^3}\right)\left(1-\frac{4}{n}\right)y^4 + \mathcal{O}\left(y^5\right) .
\end{split}
\end{equation}

Inserting the rational approximants~\eqref{HSRPADEs} into~\eqref{dotphiHphi}
leads to the corresponding slow-roll approximants based on the Pad{\'e}
approximants constructed from the Hubble expansion. In Ref.~\cite{alhugg15}
the corresponding curves in the state space ${\bf S}$ on the scalar field
subset $\Omega_m=0$ are given for $n=1$ in Fig. 10 and the associated
relative errors at the end of inflation at $q=0$ are given in Table 4. Next
we will compare some of the simplest, and thereby the most easily applicable,
approximations for each of the above schemes.

%-----------------------------------------------------------------------------
\subsection{Approximant comparisons}\label{sec:comp}
%-----------------------------------------------------------------------------

As seen from the above discussion, series expansion approximants and
slow-roll approximation schemes lead to ambiguities, i.e., different possible
types of approximations. This is to be contrasted with the center manifold
analysis, which yield a systematic scheme for obtaining a unique sequence of
converging approximations for the attractor solution. On the other hand, the
strength of slow-roll approximation schemes is that they can be applied to
any potential and problem that satisfies the slow-roll conditions; it remains
to be seen to what extent this can be accomplished with other methods like
center manifold analysis, an issue we will return to elsewhere.

%Another set of issues are accuracy, simplicity, and applicability of the
%approximations in other contexts. Simplicity is here tied to applicability,
%i.e., how `easy' it is to use the approximations in other contexts, e.g., for
%how to approximately compute the number of $e$-folds in terms of a physical
%observables and to use the approximations in cosmological perturbations.

% In this context it is of interest to point out that when it comes to slow-roll
% approximations it is usually the zeroth order slow-roll Hubble expansion,
% which as shown above is a quite bad approximation, that is used, and one
% might question if this approximation is `good enough' for precision
% cosmology.

\begin{figure}[ht!]
\begin{center}
\subfigure[$n=2$.]{\label{fig:CompPhi4}
\includegraphics[width=0.45\textwidth]{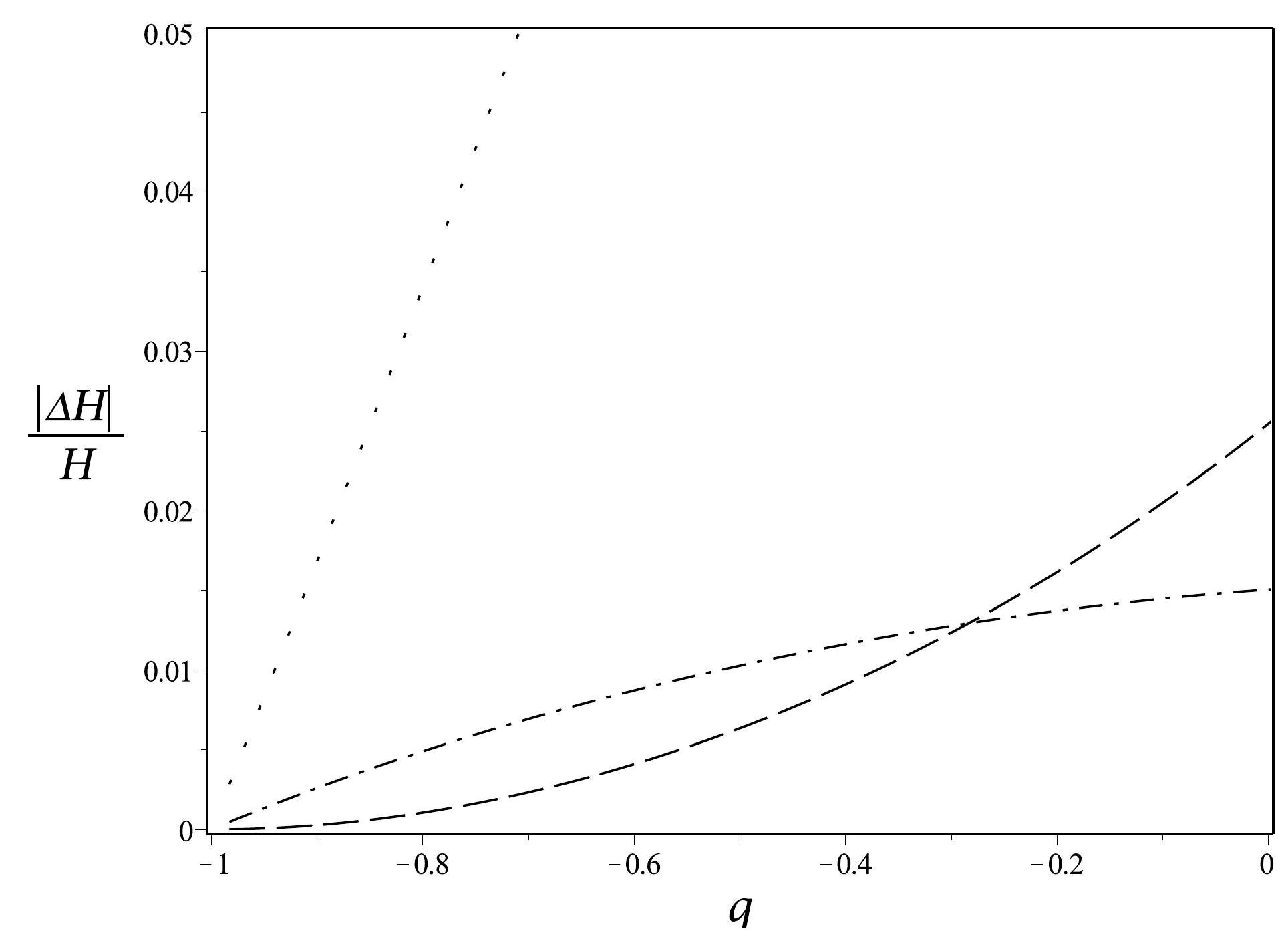}
}\quad%
\subfigure[$n=3$.]{\label{fig:CompPhi6}
\includegraphics[width=0.45\textwidth]{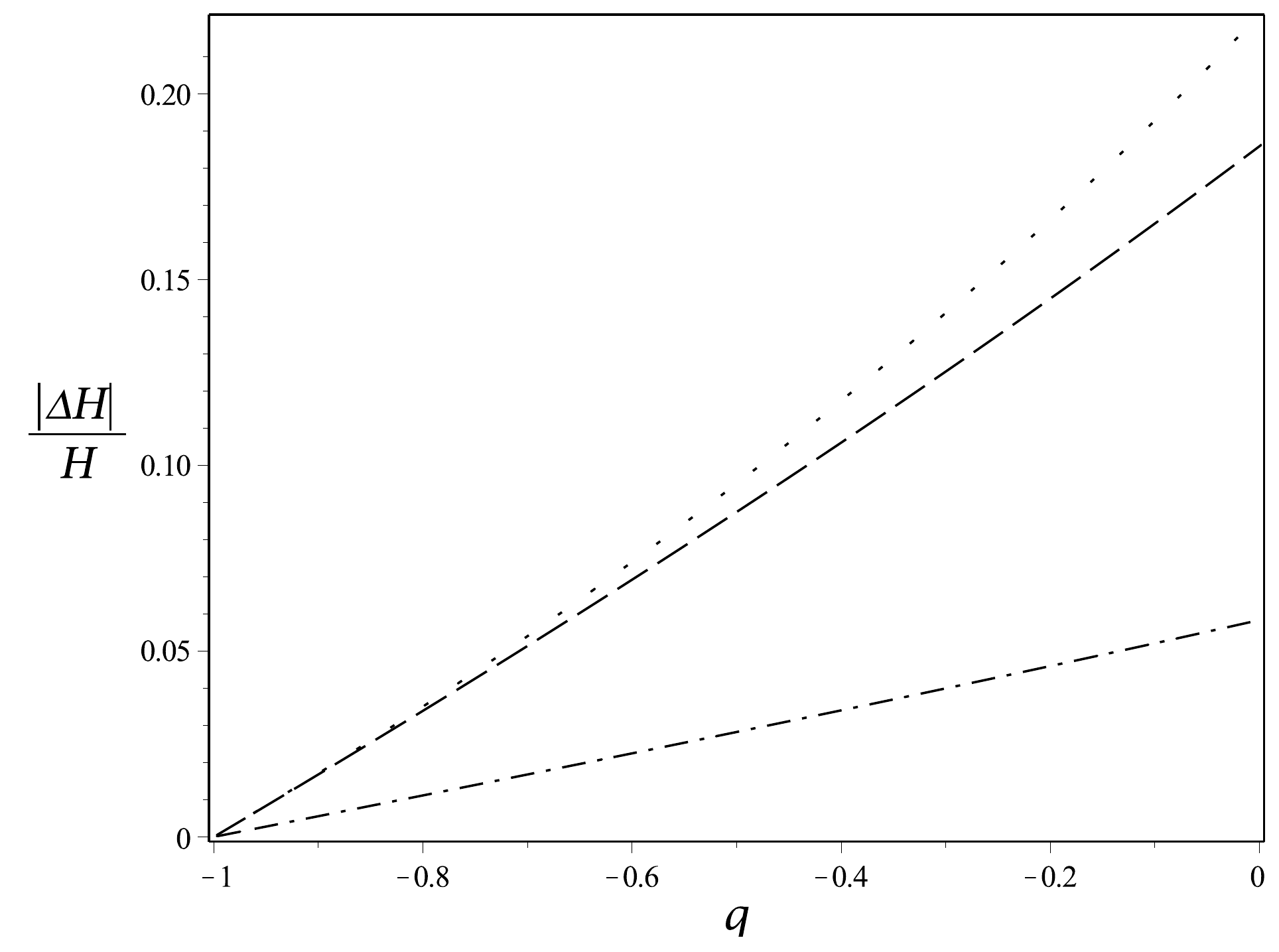}
}
\subfigure[$n=5$.]{\label{fig:CompPhi10}
\includegraphics[width=0.45\textwidth]{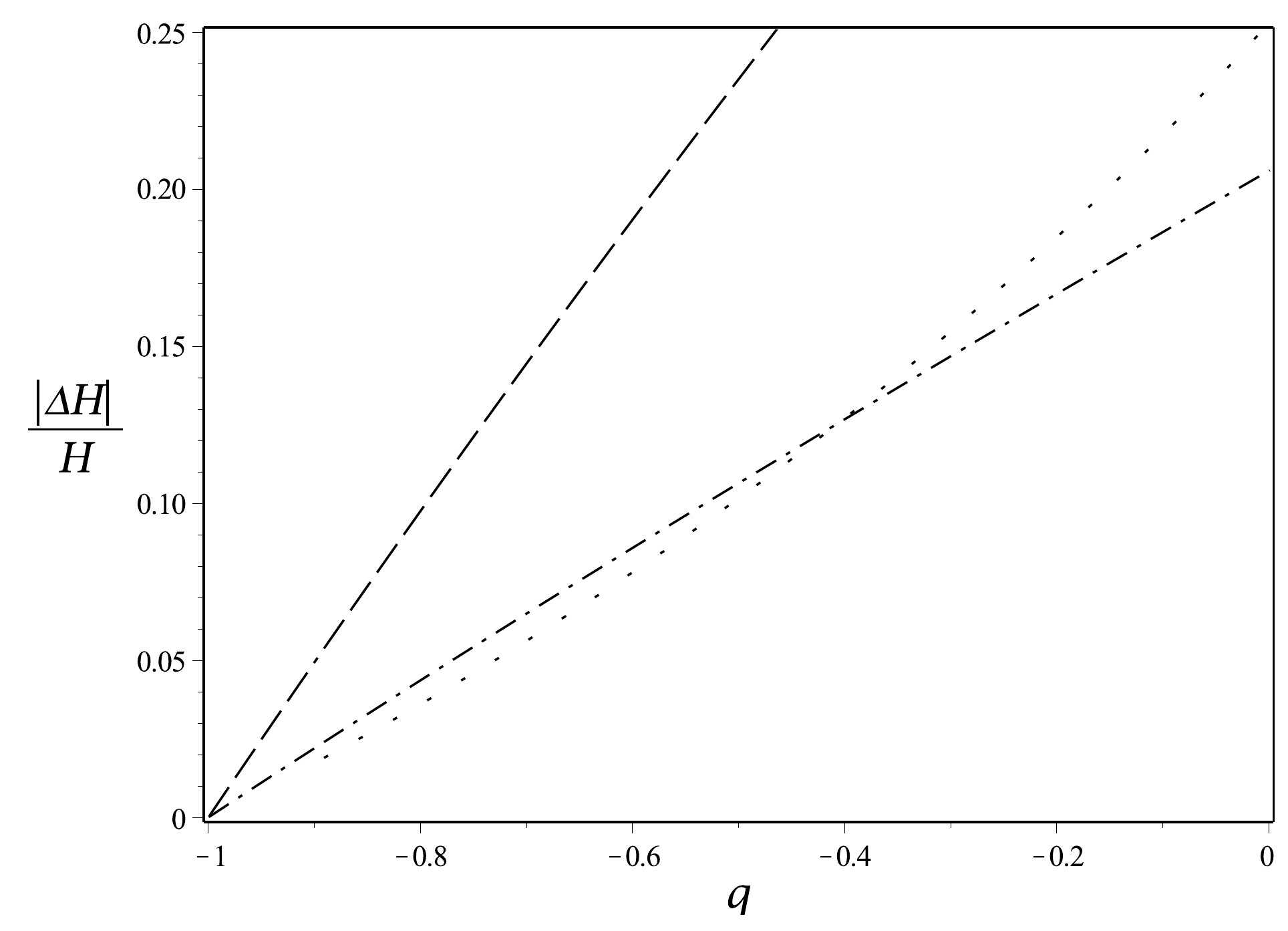}
}\quad%
\subfigure[$n=6$.]{\label{fig:CompPhi12}
\includegraphics[width=0.45\textwidth]{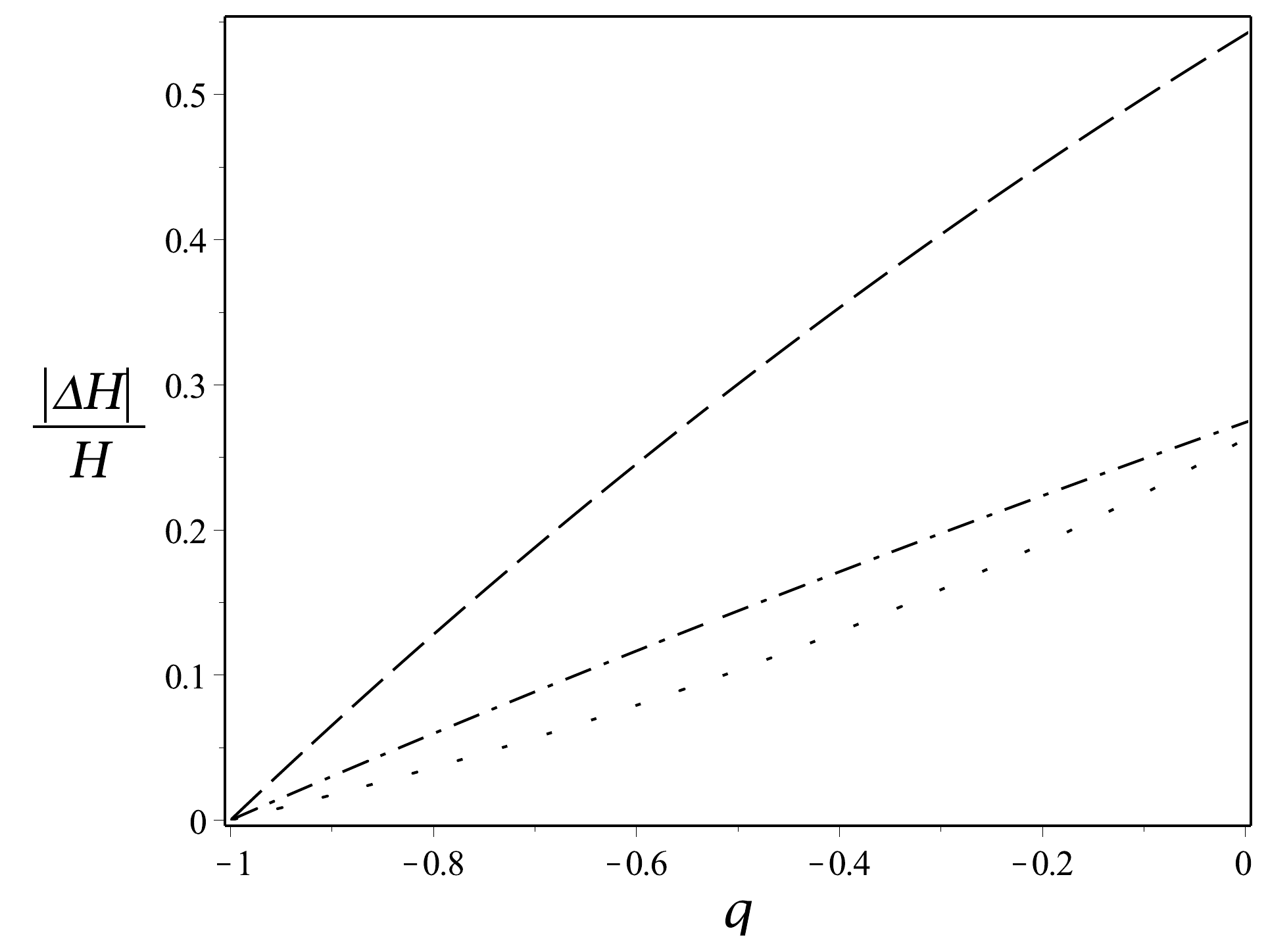}
}
\end{center}
\caption{Relative Hubble errors $\frac{|\Delta H|}{H}$  as functions of $q$ for $n=2,3,5,6$
for the slow-roll approximation (dashed curves),
the center manifold Pad{\'e} approximants $[1/1]_\theta$ (dash-dotted curves), and
the $[1/1]_{\Sigma_\dagger}$ Pad{\'e} approximants (space-dotted curves)
}%
\label{fig:ErrorsComparisons}
\end{figure}

%As it will be seen afterwards when computing the number of $e$-folds,
%we will only need a subset of the full set of approximations for the series
%and slow-roll expansions.
To keep the discussion reasonably short, we here restrict it to comparing
only some of the fairly simple approximations, namely: (i) the slow-roll
approximation given by $\Sigma_\dagger \approx -\frac{n}{3}X^{n-1}\tilde{T}$,
(ii) the center manifold Pad{\'e} approximant $\theta \approx -
\frac{\sqrt{n}}{3}\tilde{T}[1/1]_{f(x)}$ given in eq.~\eqref{CM_Pade}, which
we will refer to as $[1/1]_\theta$, and (iii) the series expansion Pad{\'e}
approximant for $\Sigma_\dagger$ given in eq.~\eqref{SigmaXPade}, which we
will refer to as $[1/1]_{\Sigma_\dagger}$. The accuracy of the different
approximations for the attractor solution, which is computed numerically, is
shown in Figure~\ref{fig:ErrorsComparisons}, while the relative errors at
$q=0$ are given in Table~\ref{tab:Comparisons}.\footnote{The relative errors
are given in terms of the Hubble variable $H$ by $|\Delta H|/H =
\frac{|H_{num} - H_{approx}|}{H_{num}} = \left|1 - \left(\frac{T_{num}(1 -
T_{approx})}{T_{approx}(1 - T_{num})}\right)^n\right|$, where the subscript
$num$ stands for the numerically computed attractor solution while the
subscript $approx$ stands for the approximant.}
\begin{table}[ht!]
\begin{center}
\begin{tabular}{|c|c|c|c|c|c|c|c|c|c|c|c|c|}  \hline
                            &             &             &             &             &            &  	   &             &	       &             &             & 		&  	 	 \\ [-2ex]
$n$     &  $1$        &     $2$     &     $3$     &     $4$     &     $5$    &     $6$     &     $7$     &       $8$   &     $9$     &     $10$    & 	$15$ 	&  $20$ 	 \\  [1ex]\hline
                            &             &             &             &             &            &  	   &             &  	       &             &             & 		& 		 \\ [-2ex]
Slow-Roll                 & $14$\%      &     $3$\%   &     $19$\%  &     $33$\%  &     $44$\% &     $54$\%  &      $62$\% &     $69$\%  &      $75$\% &    $79$\%   &  $92$\%	&  $97$\%        \\ [1ex]\hline
			    &             &             &             &             &            &		   & 	         &             &             &             & 		& 		 \\ [-2ex]
$[1/1]_{\theta}$           &   $7$\%     &    $2$\%    &    $6$\%    &    $13$\%    &     $20$\%    &        $27$\%     &       $34$\%        &    $40$\%   &   $45$\%   &   $50$\%    &  $69$\%  &  $81$\%        \\ [1ex]\hline
                            &             &             &             &              &               &                   &                     &             &            &             &          &           \\ [-2ex]                                  &             &             &            &              &               &               \\ [-2ex]
$[1/1]_{\Sigma_{\dagger}}$ &  $14$\%  &  $19$\%  & $22$\%  & $24$\% &  $25$\% & $26$\% & $27$\% &  $27.8$\% &  $28.3$\% &  $29$\% &   $30$\% &  $31$\%  \\ [1ex]\hline
\end{tabular}
\end{center}
\caption{Relative Hubble errors $\frac{|\Delta H|}{H}$ at $q=0$ for the
slow-roll $\Sigma_\dagger \approx -\frac{n}{3}X^{n-1}\tilde{T}$, and the
$[1/1]_\theta$ and $[1/1]_{\Sigma_\dagger}$ Pad{\'e} approximants.}
\label{tab:Comparisons}
\end{table}

All the presently discussed approximations, including the slow-roll
approximation $\Sigma_\dagger \approx -\frac{n}{3}X^{n-1}\tilde{T}$, have
quite small errors when $q\approx -1$, which is a reflection of that they are
all tangential to the attractor solution when $T\rightarrow 0$. For larger
$q$ the slow-roll approximation give fairly good results for small $n$ while
it becomes less competitive for steep potentials with large $n$ (recall that
$\phi \rightarrow \pm \infty$ in the de Sitter fixed point limit), as can be
expected. However, for small $n$ the slow-roll approximation lead to
fluctuations in accuracy when $q$ is no longer close to $-1$, indicating that
it sometimes might be rather hard to predict how good the slow-roll
approximation actually is going to be in a given situation.

The above suggests that the slow-roll approximation, as well as its more
accurate but more complicated higher expansion slow-roll based approximants,
should be used with care when it comes to precision cosmology. It should also
be pointed out that all approximations get worse for large $n$, hinting at
that steep potentials might pose a challenge for approximative methods
(although the physical viability of such models might be questioned in view
of recent observational data). However, the $[1/1]_{\Sigma_\dagger}$ Pad{\'e}
approximant does not get worse so fast as the other approximations and for
$n\geq6$ it is the best approximation of the three. This suggests that a good
analytic description, pertinent in the context of precision cosmology, might
require the development of a variety of complementary approximation methods
that can be applied to a range of different problems; we emphasize that the
present problem is just an illustrative stepping stone for such developments.

%%%%%%%%%%%%%%%%%%%%%%%%%%%%%%%%%%%%%%%%%%%%%%%%%%%%%%%%%%%%%%%%%%%%%%%%%%%%%%
\section{Concluding remarks}\label{sec:concl}
%%%%%%%%%%%%%%%%%%%%%%%%%%%%%%%%%%%%%%%%%%%%%%%%%%%%%%%%%%%%%%%%%%%%%%%%%%%%%%

In this paper we have used the inherent physical scales associated with the
problem at hand to produce new variables. In particular we obtained a regular
dynamical system~\eqref{dynsys} on a compact state space $\bar{\bf S}$. This
allowed us to give a pictorial description of the entire solution space and
moreover give proofs concerning the solutions' asymptotic properties, which
in turn were tied to physically important features such as asymptotic
manifest self-similarity and breaking thereof. In addition we showed that the
so-called attractor solution is just the unstable center manifold of a de
Sitter state on the unphysical past boundary subset on $\bar{\bf S}$. We also
found it convenient to introduce several auxiliary dynamical systems, adapted
to the physical properties associated with asymptotic past dynamics, since
this allowed us to efficiently compute approximations for the attractor
solution by means of several complementary approximation schemes. The common
feature of these schemes is that to lowest order toward the de Sitter point
(apart from those approximations like the lowest order Hubble expansion
slow-roll approximation, which just give vertical lines at $q=-1$), from
which the attractor solution originate, the approximations yield the tangent
space of the center manifold. As a consequence they all lead to that the
variable $\tilde{T}$ along the attractor solution in the limit of small
$\tilde{T}$ is governed by the approximation
\begin{equation}
\frac{d\tilde{T}}{d{\tau}} = \frac{n}{3}\,\tilde{T}^3.
\end{equation}
This, in combination with the definition $\tilde{T} = c H^{-\frac{1}{n}}$ and
that $\tau$ can be identified as $\ln a$, yields that when the scale factor
$a\rightarrow 0$ then
\begin{equation}
H \propto \big[\ln (a_0/a)\big]^{\frac{n}{2}} \rightarrow \infty,
\end{equation}
which is to be contrasted with the exact de Sitter solution for which $H =
\mathrm{constant}$.

Finally, we would like to point out that in principle the methods we have
introduced have a broader range of applicability than the presently studied
models. For example, the use of averaging techniques should be
applicable to the dynamics at late times for solutions that approaches a
local minimum of a scalar field potential of the form
\begin{equation}
V \propto \phi^{2n}(1 + W(\phi)) \quad \text{where} \quad \lim_{\phi \rightarrow 0}W =0,
\end{equation}
where we for simplicity have translated the minimum of the potential so that
it occurs at $\phi=0$, and where $W$ obeys suitable differentiability
conditions when $\phi \rightarrow 0$. One might question the physical
relevance of these potentials at very late times, but it should be pointed
out that asymptotic dynamics can be used as pieces to provide approximations
for global dynamics, as described in~\cite{alhugg15}.

%%%%%%%%%%%%%%%%%%%%%%%%%%%%%%%%%%%%%%%%%%%%%%%%%%%%%%%%%%%%%%%%%%%%%%%%
\subsection*{Acknowledgments}
AA is supported by the projects CERN/FP/123609/2011, EXCL/MAT-GEO/0222/2012,
and CAMGSD, Instituto Superior T{\'e}cnico by FCT/Portugal through UID/MAT/04459/2013,
and the FCT grant SFRH/BPD/85194/2012. Furthermore, AA also thanks the
Department of Physics at Karlstad University, Sweden, for kind hospitality.
JH is supported by the DFG collaborative research center SFB647 Space, Time,
Matter. CU would like to thank the Center for Mathematical Analysis, Geometry
and Dynamical Systems at the Technical University of Lisbon and the Institut
f\"ur Mathematik at Freie Universit\"at in Berlin for kind hospitality.
Finally, it is a pleasure to thank Prof. B. Fiedler and Dr. S. Liebscher for
useful suggestions.

\label{bibbegin}


\begin{thebibliography}{99}

\bibitem{tur83} M.~S.~Turner.
\newblock Coherent scalar-field oscillations in an expanding universe.
\newblock Phys. Rev.\ {\bf D28} 1243 (1983).% 1243-1247
\newblock DOI: 10.1103/PhysRevD.28.1243.

\bibitem{muk05} V.~Mukhanov.
\newblock {\em Physical foundations of Cosmology}.
\newblock Cambridge University Press, Cambridge, (2005).

\bibitem{ren07} A.~D.~Rendall.
\newblock Late-time oscillatory behaviour for self-gravitating scalar fields.
\newblock Class.\ Quant.\ Grav.\ {\bf 24} 667 (2007).% 667-678
\newblock DOI: 10.1088/0264-9381/24/3/010.

\bibitem{macpic00} A.~de la Macorra and G.~Piccinelli.
\newblock Cosmological evolution of general scalar fields and quintessence.
\newblock Phys.\ Rev.\ {\bf  D61}, 123503 (2000).
\newblock DOI: 10.1103/PhysRevD.61.123503.

\bibitem{giamir10} R.~Giambo and J.~Miritzis.
\newblock Energy exchange for homogeneous and isotropic universes with a scalar field
coupled to matter.
\newblock Class.\ Quant.\ Grav.\  {\bf 27} 095003 (2010).
\newblock DOI: 10.1088/0264-9381/27/9/095003.


\bibitem{col03} A.~A.~Coley.
\newblock {\em Dynamical systems and cosmology}.
\newblock Kluwer Academic Publishers, Dordrecht, (2003).

\bibitem{beyesc13} F.~Beyer and L.~Escobar.
\newblock Graceful exit from inflation for minimally coupled Bianchi A scalar field
models.
\newblock Class.\ Quant.\ Grav.\ {\bf 30} 195020 (2013).
\newblock DOI: 10.1088/0264-9381/30/19/195020

% \bibitem{fadetal14} C.~R.~Fadragas, G.~Leon and E.~N.~Saridakis.
% \newblock Dynamical analysis of anisotropic scalar-field cosmologies for a wide range
% of potentials.
% \newblock  Class.\ Quant.\ Grav.\ {\bf 31} 075018 (2014).
% \newblock DOI: 10.1088/0264-9381/31/7/075018.

\bibitem{fadetal14} G.~Leon and C.~R.~Fadragas.
\newblock Cosmological Dynamical Systems: And their Applications.
\newblock LAP LAMBERT Academic Publishing (2012).

\bibitem{ugg13b} C.~Uggla.	
\newblock Recent developments concerning generic spacelike singularities.
\newblock Gen.\ Rel.\ Grav.\ {\bf 45} 1669 (2013). % 1669�1710
\newblock DOI: 10.1007/s10714-013-1556-3.

\bibitem{reyure10} M.~J.~Reyes-Ibarra and L.~A.~Ure\~na-L\'{o}pez.
\newblock Attractor dynamics of inflationary monomial potentials.
\newblock AIP Conference Proceedings {\bf 1256}, 293 (2010).
\newblock DOI: 10.1063/1.3473869.

\bibitem{uggetal95} C.~Uggla, R.~T.~Jantzen, and K.~Rosquist.
\newblock Exact hypersurface homogeneous solutions in cosmology and astrophysics.
\newblock Phys.~Rev. {\bf D51} 5522 (1995). %5522-5557
\newblock DOI: 10.1103/PhysRevD.51.5522

\bibitem{alhugg15} A.~Alho and C. Uggla.
\newblock Global dynamics and inflationary center manifold and slow-roll approximants.
\newblock J. Math. Phys. {\bf 56} 012502 (2015).% -
\newblock DOI: 10.1063/1.4906081

\bibitem{beletal85a} V.~A.~Belinski\v{\i}, L.~P.~Grishchuk, I.~M.~Khalatnikov
    and Y.~B.~Zeldovich.
\newblock Inflationary stages in cosmological models with a scalar field.
\newblock Sov.\ Phys.\ JETP\ {\bf 62} 195 (1985).

\bibitem{kistim08} V.~V.~Kiselev and S.~A.~Timofeev.
\newblock Quasiattractor dynamics of $\lambda\phi^4$-inflation.
\newblock  arXiv:0801.2453 (2008).

\bibitem{lin83} A.~D.~Linde.
\newblock Chaotic inflation.
\newblock Phys.\ Let. {\bf B 129} 177 (1983).

\bibitem{cra91} J.~D.~Crawford.
\newblock Introduction to bifurcation theory.
\newblock Rev.\ Mod.\ Phys. {\bf 63} 991 (1991). %991-1038

\bibitem{waiell97} J.~Wainwright and G.~F.~R.~Ellis.
\newblock {\em Dynamical systems in cosmology}.
\newblock {C}ambridge {U}niversity {P}ress, Cambridge, (1997).

\bibitem{ren02} A.~D.~Rendall.
\newblock Cosmological models and centre manifold theory.
\newblock Gen. Rel. Grav.\ {\bf 34} 1277 (2002).% 1277-1294
\newblock DOI: 10.1023/A:1019734703162

\bibitem{waietal99} J.~Wainwright, M.~J.~Hancock and C.~Uggla.
\newblock Asymptotic self-similarity breaking at late times in cosmology.
\newblock Class.\ Quant.\ Grav.\ {\bf 16} 2577 (1999).% 2577-2598
\newblock DOI: 10.1088/0264-9381/16/8/302.

\bibitem{def75} L.~Defrise-Carter.
\newblock Conformal Groups and Conformally Equivalent Isometry Groups.
\newblock Commun.\ math.\ Phys.\ {\bf 40} 273 (1975). %273-282

\bibitem{ear74} D.~M.~Eardley.
\newblock Self-similar Spacetimes: Geometery and Dynamics.
\newblock Commun.\ math.\ Phys.\ {\bf 37} 287 (1974). %287-309

\bibitem{remcar13} G.~N.~Remmen and S.~M.~Carroll.
\newblock Attractor solutions in scalar-field cosmology.
\newblock Phys.\ Rev.\ {\bf D88}  (2013).% -
\newblock DOI: 10.1103/PhysRevD.88.083518

\bibitem{remcar14} G.~N.~Remmen and S.~M.~Carroll.
\newblock How many e-folds should we expect from high-scale inflation?
\newblock Phys.\ Rev.\ {\bf D90}  (2014).% -
\newblock DOI: 10.1103/PhysRevD.90.063517

\bibitem{corslo14} A.~Corichi and D.~Sloan.
\newblock Inflationary Attractors and their Measures.
\newblock Class.\ Quant.\ Grav.\ {\bf 31} 062001 (2014).
\newblock DOI: 10.1088/0264-9381/31/6/062001

\bibitem{lidetal94} A.~R.~Liddle, P.~Parsons and J.~D.~Barrow.
\newblock Formalizing the slow-roll approximation in inflation.
\newblock Phys.\ Rev.\ {\bf D50} 7222 (1994).% 7222-2232
\newblock DOI: 10.1103/PhysRevD.50.7222

\bibitem{car81} J.~Carr.
\newblock {\em Applications of center manifold theory}.
\newblock Springer Verlag, New York, 1981.

\bibitem{sinetal03} J.~J.~Sinou, F.~Thouverez and L.~Jezequel.
\newblock Extension of the center manifold approach, using rational fractional
approximants, applied to non-linear stability analysis.
\newblock Nonlinear Dynamics\ {\bf 33} 267  (2003).% 267-282
\newblock DOI: 10.1023/A:1026060404109

\bibitem{bohetal12} C.~G.~B{\"o}hmer, N.~Chan and R.~Lazkoz.
\newblock Dynamics of dark energy models and centre manifolds.
\newblock Phys.\ Lett.\ B, {\bf 714}, 11 (2012). %11-17
\newblock DOI: 10.1016/j.physletb.2012.06.064.

\bibitem{bak75} G.~A.~Baker.
\newblock {\em Essentials of Pad\'e Approximants}.
\newblock {A}cademic {P}ress, New York, (1975).

\bibitem{kal02} J.~Kallrath.
\newblock {\em On Rational Function Techniques and Pad{\'e} Approximants. An
Overview.}
\newblock (2002).

\bibitem{salbon90} D.~S.~Salopek and J.~R.~Bond.
\newblock Nonlinear evolution of long-wavelength metric fluctuations in
inflationary models.
\newblock Phys.\ Rev.\ {\bf D42} 3936 (1990).

%\bibitem{chi87} C.~Chicone.
%\newblock The monotonicity of the period function for planar Hamiltonian
%vector fields.
%\newblock J. Diff. Eq. {\bf 69}, 310 (1987). % 310-321

\bibitem{SanVerMu}J.~A.~Sanders, F.~Verhulst, J.~Murdock
\newblock{\em Averaging Methods in Nonlinear Dynamical Systems}
\newblock Applied Mathematical Sciences 59, Springer (2007).

\bibitem{GuckHo} J.~Guckenheimer, P.~Holmes.
\newblock{\em Nonlinear Oscillations, Dynamical Systems, and Bifurcations of Vector Fields}.
\newblock Springer, Applied Mathematical Sciences, 42 (2000).

\bibitem{CoLev} E.~A.~Coddington, N.~Levinson.
\newblock{\em Theory of Differential Equations}
\newblock McGraw-Hill, New Yorck (1955).

%%%%%%%%%%%%%%%%%%%%%%%%%%%



\end{thebibliography}
\end{document}